\documentclass[
aps,
pra,
10pt,
twocolumn,
]{revtex4-2}

\usepackage[english]{babel}
\usepackage[utf8]{inputenc}
\usepackage{amsmath,amsbsy,amssymb,amsthm,amsfonts}
\usepackage{dsfont}
\usepackage{xcolor}
\usepackage{bm}
\usepackage{physics}
\usepackage{upgreek}
\usepackage{graphicx}
\usepackage{float}
\usepackage{nicefrac}
\usepackage[version=4]{mhchem}
\usepackage{hyperref}
\hypersetup{
colorlinks=true,
linkcolor=blue,
filecolor=magenta,
urlcolor=cyan
}
\usepackage{multirow}
\newtheorem{task}{Task}

\newtheorem{problem}{Problem}

\newcommand{\da}[1]{#1^{\dagger}}

\newcommand{\twoq}{2QG}
\newcommand{\Inf}{$\mathcal{I}$}

\newcommand{\MS}{M\o{}lmer--S\o{}rensen}
\newcommand{\mmc}{AM-MMC}

\newcommand{\ps}{\textit{phase-space}}
\newcommand{\sota}{SotA}

\newcommand{\e}{\mathrm{e}}
\newcommand{\ii}{\mathrm{i}}
\newcommand{\sdf}{\textit{spin-dependent-force}}

\newcommand{\schr}{Schr\"{o}dinger}
\newcommand{\com}{{CoM}}

\newcommand{\qme}{{QME}}
\newcommand{\ito}{It\^o}

\newcommand{\ael}[3]{$^{#1} \mathrm{#2}_{\nicefrac{#3}{2}}$} 
\newcommand{\rad}{\rm s^{-1}}
\newcommand{\us}{$\upmu$s}
\begin{document}
\title{Closed-loop control for two-qubit gates with trapped ions}

\author{Eduardo J.~P\'aez\href{https://orcid.org/0000-0001-8342-7090}{\includegraphics[scale=0.05]{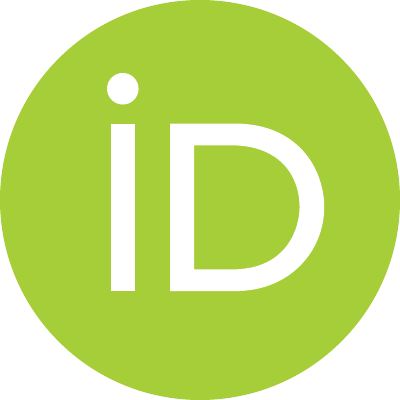}}}
\altaffiliation{These authors contributed equally}
\email{eduardo.paez1@ucalgary.ca}
\affiliation{Institute for Quantum Science and Technology, University of Calgary, Alberta T2N~1N4, Canada}

\author{Seyed Shakib Vedaie\href{https://orcid.org/0000-0001-5025-8585}{\includegraphics[scale=0.05]{figures/orcidid.pdf}}}
\altaffiliation{These authors contributed equally}
\email{seyedshakib.vedaie@alumni.ucalgary.ca}
\affiliation{Institute for Quantum Science and Technology, University of Calgary, Alberta T2N~1N4, Canada}

\author{Barry C.~Sanders\href{https://orcid.org/0000-0002-8326-8912}{\includegraphics[scale=0.05]{figures/orcidid.pdf}}}
\email{sandersb@ucalgary.ca}
\affiliation{Institute for Quantum Science and Technology, University of Calgary, Alberta T2N~1N4, Canada}

\date{\today}
\begin{abstract}
State-of-the-art two-qubit gates with trapped ions employ open-loop control that
rely on simplified models to precompute control sequences.
Our aim is to introduce closed-loop control for two-qubit gates to correct disturbances as they occur during the gate implementation.
We introduce a spectator ion into the ion chain used for quantum logic processing, where it couples with the other ions through collective motional modes. The spectator ion's position
is continuously monitored by driving dipole transitions and detecting the resultant fluorescence.
We show that incorporating a spectator ion is feasible for linear Paul trap implementations and is expected to reduce the two-qubit gate Bell-state preparation
infidelity by an order of magnitude with the deleterious effects of position monitoring being negligible compared 
to the thermal effects that exist in the system, even in the absence of the spectator ions.
Mathematically, we describe driven ion-trap dynamics, including the spectator ion, by a stochastic quantum master 
equation involving the amplitude-modulation multimode-motional coupling gate, motional drift, thermal effects, recoil from photon scattering,
spontaneous decay, and light shift.
Our on-the-fly control method employs reinforcement learning with 
the reward function based on the actual geometric phase of the spectator ion. A key advantage of our approach is that we introduce a control method that involves 
`learning' and correcting disturbances happening in the trap on-the-fly, thus 
achieving high-fidelity gates.
Our approach will lead to a significantly higher two-qubit gate fidelity at a reduced calibration overhead owing to the small parameter drift in the control system. 
\end{abstract}

\maketitle

\section{Introduction}\label{sec:intr}
High-quality or low-infidelity two-qubit gate (\twoq) implementations are vital for scalable quantum information processing with trapped ions~\cite{EDN+21, SBK+21, ACC+21, CNE+23, BML24, RAA+26}.
Currently, the state-of-the-art control procedures~\cite{CDM+14, WBD+19, GTL+16, KWF+23} in ion-trap platforms do not yet deliver reliable control 
policies to implement \twoq\ that meet the fault-tolerant threshold with an increasing number of ions~\cite{SBK+21, ACC+21, CTS+21, BHL+16, KWF+23, SNO+24, EG25}.
We develop a control scheme by extending a 
previously validated model~\cite{VPS23} that allows us to compensate for errors and decoherence in real time. Our scheme incorporates position 
monitoring of a spectator ion~\cite{CAP+21_prl} and learning to devise robust control policies against motional drift, heating, and other 
decoherence sources that are typical in ion-trap-based gate implementations~\cite{LH24, WHT20}.

The state-of-the-art (\sota) methods for implementing gates in ion traps are based on open-loop control schemes~\cite{BCM+19, CDM+14, WBD+19, GTL+16, BHL+16, SBK+21}. 
The gate-design policy is computed based on the presumptions of closed-system ion-chain dynamics, leading to models that are not fully reliable for precise quantum control.
Additionally, open-loop control policies are intrinsically blind to the numerous noise sources interacting with the system during gate implementations~\cite{BML24, VPS23, MEH+20}.

Recently, global optimisation algorithms have been used to find robust and feasible 2QG control policies under laser-power and gate-duration constraints~\cite{VPS23}.
Nonetheless, once the Hamiltonian parameters change, either drifting (e.g.,~motional frequency drift) or fluctuating (e.g.,~motional heating), 
beyond the robustness target, the search algorithm must be performed again to find new solutions, which are computationally expensive and time-consuming.

The intrinsic limitations of the open-loop control schemes discussed above have motivated the search for alternative approaches based on real-time closed-loop control~\cite{VPS23, BCM+19, VDP+23}. In this context, the rich theory of continuous monitoring of quantum systems~\cite{JS06}, as described by the quantum stochastic master equation~\cite{GC84,BHJ07}, provides an essential foundation for further developments.

Continuous monitoring in linear Paul traps was first implemented to study light interference of a single ion and its 
mirror image~\cite{ERSB01}. In that work, the authors showed that, by placing a single~$^{138}$Ba$^{+}$ with a back mirror, 
the position of the ion can be continuously monitored. They were able to resolve the ion's position with
a resolution of 1.7~nm, which is more precise than the spread of the ground state wave packet of 7~nm. This 
idea was further developed by employing closed-loop control to cool a 
single ion below its Doppler limit~\cite{BRW+06}.
The core idea lied in gaining knowledge of the ion's position from 
its scattered light, followed by feeding back that signal to the ion-trap electrodes 
with the proper phase and gain, thereby generating a dragging force on the velocity of the ions.

In this work we develop a closed-loop control scheme based on continuous position 
monitoring of a spectator ion involving a learning agent, which we abbreviate as CLPM-scheme: closed-loop learning-control position monitoring. Our CLPM-scheme 
allows us to compensate for the aforementioned disturbances in real time. We extend the comprehensive model of open-system 
dynamics for trapped ions~\cite{VPS23}, based on quantum trajectory theory (QTT)~\cite{MKS+19} to incorporate the continuous monitoring of the spectator ion. The extended model serves as a platform for estimating the diffusive effects of continuous monitoring during \twoq\ implementation. 
We demonstrate that continuous position monitoring of an ancillary ion in a three-ion chain is technically feasible and has an impact on the \twoq\ infidelity~$\mathcal{I}$~\cite{CDM+14} of one order of magnitude lower than the effects caused by motional heating at similar rates. Our findings indicate that the rate of photodetection can reliably provide sufficient information about the position and momentum of ions with a latency compatible with the \twoq\ cycle time on trapped-ion platforms. Furthermore, we demonstrate 
that an agent employing machine learning can devise robust control policies against motional drift, heating, and light shifts.

The novelty of our work is twofold. First, we synthesise novel techniques in real-time quantum measurement for position monitoring of an ancillary ion during \twoq\ implementation.
Second, we incorporate reinforcement learning into a closed-loop system to devise robust control policies.
The developed control scheme is important for designing \twoq\ as it opens the way for to mitigate disturbances 
and decoherence in real time, which pose major obstacles to scalable fault-tolerant quantum computing on ion-trap platforms.
Our developed technique reduces the control complexity owing to parameter variations over time and could also be applicable to other quantum computing platforms.

\section{Background}\label{sec:background}
We now provide the salient background for our work. We elaborate on the relevant information for quantum computing based on trapped ions, the control tasks leading to \twoq implementations, and ion-position monitoring.

\subsection{Trapped ions for quantum computing} \label{sec:back:trappedionqcomp}
We elaborate on the key aspects of ion dynamics in linear Paul traps.
First, we describe the generation of spin-dependent motional coupling in the proposed device. Second, we explain the role of motional dynamics on the accumulation of geometric phases that lead to 
entanglement. Finally, we discuss the use of ancillary ions to enhance quantum information processing.

\subsubsection{Spin-dependent motional coupling}

Entanglement generation relies on exploiting the strong interactions between different degrees of freedom in quantum systems.
In trapped-ion dynamics, the Coulomb interaction enables the internal electronic states of the trapped ions to couple with the collective motion of the confined crystal.
Because the Coulomb interaction is strong and long-ranged~\cite{DDM+23}, each ion responds concomitantly to photon absorption events that occur anywhere in the crystal.
By judiciously choosing the Hamiltonian parameters and the duration of the interaction, the target degrees of 
freedom can reach a maximally entangled correlation state.

The energy exchange between the electronic states of the ions and the collective motion of the crystal is mediated by the recoil of the photon absorption. 
The coupling of laser light with the radiative moments of ions drives energy transitions by absorbing and scattering photons.
Taking the effective electromagnetic $\bm{k}$-vector of a laser impinging on an ion detuned by $\mu$ from the bare electronic transition $\omega_{\rm eg}$, and the Pauli operator $\sigma^\text{x}$, the coupling to the motion of 
the ion oscillating at $\nu$ in a given orientation $\bm{r}$ is ($\hbar\equiv1$)
\begin{equation}
\label{eq:Hj}
H_{\jmath}=\Omega_{\jmath} \sigma_{\jmath}^\text{x} \e^{\ii \bm{k}\cdot\bm{r} -\ii (\mu-\nu) t} + \text{hc},
\end{equation}
where $\Omega_{\jmath}$ is the Rabi frequency driving the $\jmath^{\rm th}$-ion and hc is the Hermitian conjugate.
The direction of the force applied to each coherent motional mode $\dot{\bm{p}}_\jmath=-\nicefrac{\partial H_{\jmath}}{\partial \bm{r}}$ reveals its dependence on the internal electronic state of the qubit.

Linear Paul traps are notable in the quantum-computing community because of their rich spin-bosonic description and highly controllable degrees of freedom~\cite{MEK00, EK04}.
In particular, one of the widely used \sota\ methods is the amplitude-modulated multimode motional coupling (\mmc) version of the \MS\ gate~\cite{DLF+16, CDM+14, WBD+19}. Such a method offers the advantages of highly controllable manipulation of electronic states using commercially available laser-coherent light and robustness against motional heating.
Additionally, the \mmc\ scheme offers a rich set of additional parameters for a more precise control of the excursion of the \ps\ trajectory.

\subsubsection{Ion-crystal motional dynamics}

\begin{figure}
	\centering
	\begin{tabular}{c}
	\includegraphics[width= 0.6 \linewidth]{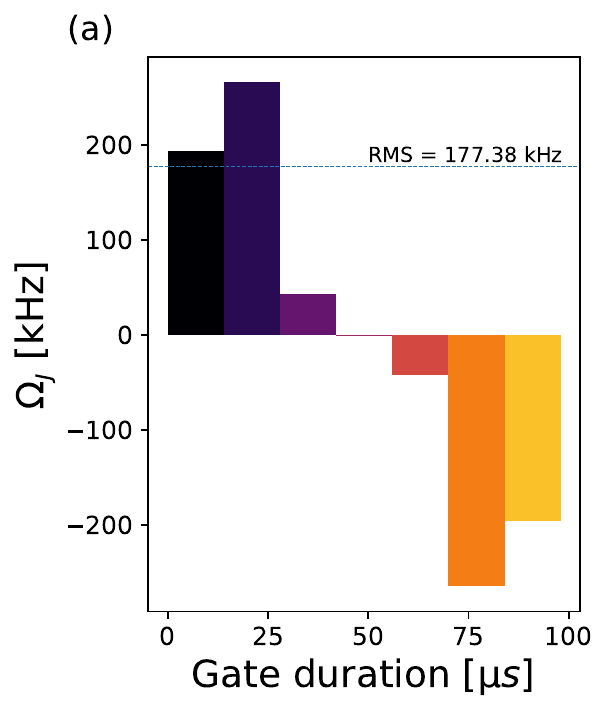}\\
	\includegraphics[width=\linewidth]{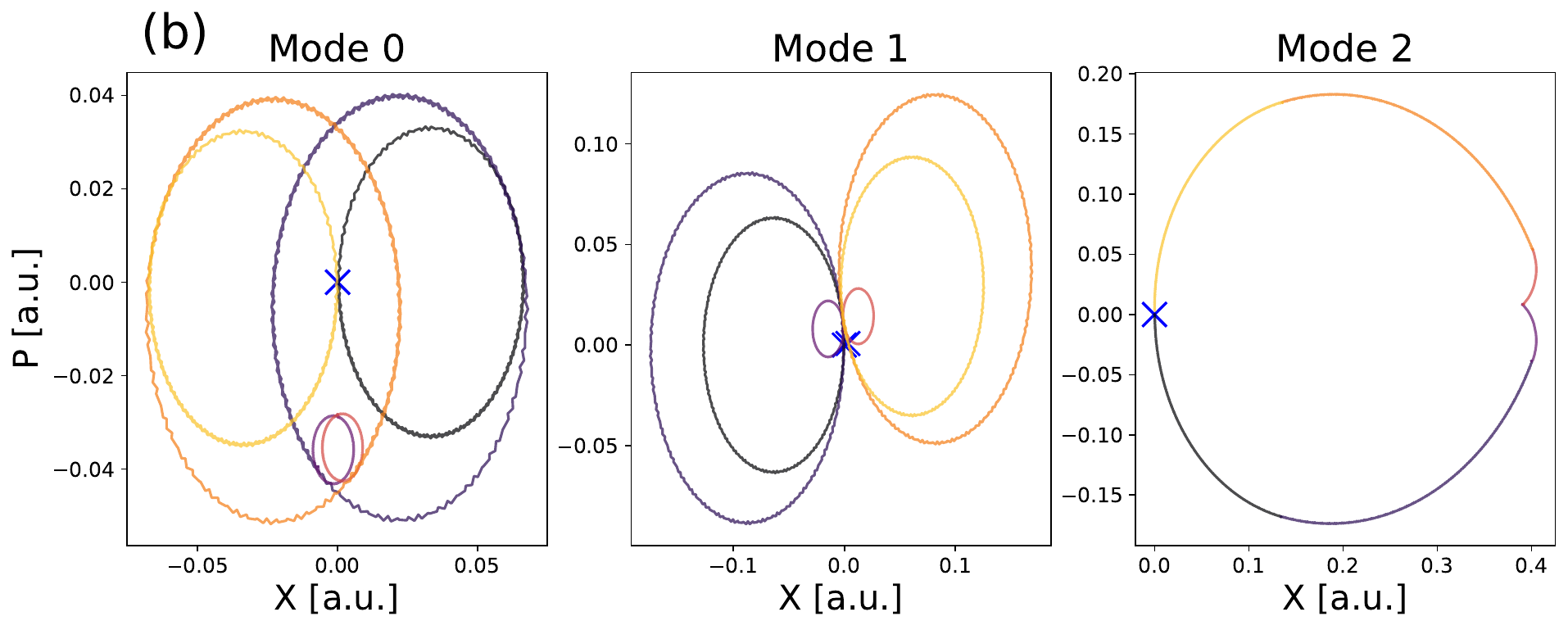}	
	\end{tabular}
	\caption[Pulse sequence designed by employing \sota\ method.]
	{(a) Pulse sequence designed by employing the \sota\ method and the corresponding \ps\ trajectories.
		The colours are arbitrary and intended to guide the eye over each piecewise constant segment. The horizontal dashed line represents the root mean square (RMS) value of the pulse sequence.
		(b) Ideal \ps\ trajectories. The quadrature components of position $X$ and momentum $P$ are given in arbitrary units (a.u.). `Mode 0' is the \com, `Mode 1' is the tilt mode, and `Mode 2' the `plucky' mode. The colour of the trajectory is associated with the piecewise-constant segment of the pulse sequence shown in (a). The blue `X' markers in (b) indicate the starting and ending points of the dynamical-phase evolution.}
	\label{fig:ideal_traj}
\end{figure}

As the $k^{\rm th}$ motional mode of the $\imath^{\rm th}$- and $\jmath^{\rm th}$-ion makes an excursion in the phase space, its evolution is described by the dynamical phase~$\alpha_{\jmath k}(t)$ and the accrued 
geometric phase of the pair of ions and $k$-motional mode, $\chi_{\imath\jmath}^{(k)}(t)$~\cite{CDM+14, VPS23, LLF+18, KWF+23}.
When the ions are illuminated with a laser detuning close to the motional frequencies, their electronic  
states interact via motional modes, thereby creating effective spin interactions. 
At a given gate duration~$\tau_\text{g}$, the right amount of $\sum_k\chi_{\imath\jmath}^{(k)}(\tau_\text{g})$ gives origin to a coherent superposition of joint electronic 
states~\cite{MC+21, Isl12}.
In Fig.~\ref{fig:ideal_traj} we illustrate a three-ion chain promoted by the \mmc\ scheme. The seven-segment piecewise constant pulse shape is illustrated in Fig.~\ref{fig:ideal_traj}(a).
The number of segments matches the number of variables to be resolved $2N+1$ such that the final state is a maximally entangled state. 
The excursions of the three motional modes in the phase space are shown in Fig.~\ref{fig:ideal_traj}(b), respectively.
The number of loops is proportional to the difference $\abs{\nu_k-\mu}$. 
Note that every motional trajectory is closed, implying that no residual correlation remains between the electronic states and the motional modes; therefore, in principle, an ideal \twoq\ implementation can be achieved.

\subsubsection{Ancillary ions for enhanced quantum information processing}
In the following paragraphs, we describe two relevant works that employ spectator or sensor ions for quantum information processing.
First, we describe the protection of qubit coherence from accidental measurements using sensor ions.
Second, we describe the implementation of noise mapping in a two-dimensional array by employing spectator ions.
Finally, we describe the sympathetic cooling technique.

A control scheme was presented to protect the coherence of the qubit space during state-destroying measurements on neighbouring ions~\cite{MK+24}. Such disturbances occur in protocols such as quantum error correction or sympathetic cooling.
This work employed sensor ions to correct for optical aberrations and an intensity probe with  $>50$~dB dynamic range. 

Another application deals with noise mapping for quantum-architecture control schemes~\cite{GGB20, GE+20}. 
Their objective is to spatially distribute the noise sensing and classical correlations by employing a subset of `sensor' qubits.
The sensors characterise the spatial variations in the processes inducing errors to determine which qubits should be actuated using information from a specific sensor.

Another important application is the treatment of the problem of cooling long ion crystals. 
A broad variety of laser-based cooling techniques utilise allowed electronic transitions, impeding the protection of the qubit space during runtime~\cite{MQ+23, WH+25}.
Sympathetic cooling is proposed as a suitable technique that uses a different atomic species to cool the chain, whereas quantum logic operations are performed on the computational ions. 
A different scheme removes the need for a different atomic species in the ion crystal using feedback cooling~\cite{SRZ05, SHR+12, ZWW+17}. This scheme employs ion-position monitoring which we will describe in \S\ref{sec:backg:ion_position}.

\subsection{Control task for two-qubit gate implementations} \label{sec:back:control_2qg}
Now we describe SotA modelling for a \twoq\ with trapped 
ions based on AM pulse-shaping to optimise multimode motional coupling for entangling ions.
To begin with, we describe the physical model for effecting this two-qubit gate via Raman transitions, the corresponding control tasks, and finally, we describe the \sota\ approach to simulating the \twoq\ with open-system dynamics.

\subsubsection{\twoq\ using Raman transitions}

Let us describe the generalised assumptions made in the \sota\ methods. 
First, the \sota\ methods for the \twoq\ design assume unitary processes; therefore, they lack physical reliability and the interplay of loss and decoherence during the \twoq\ implementation.
State-of-the-art models further assume that the internal states of each ion are treated as a two-level system (2LS) with~$\ket2$ adiabatically 
decoupled from the upper energy level with~$\Omega_\jmath(t)\leq\Omega_\text{max}$.
The laser detuning~$\mu(t)\equiv\mu$ is constant,
and the model is restricted to identical pulses for each of the two ($q$ and $r$) target ions for the \twoq\ and zero driving for the other ions.
All motional modes are in thermal equilibrium in the Lamb-Dicke regime~\cite{LMR+17}.

Accounting analytically for Rabi-dependent energy shifts, such as the Autler--Townes and Bloch--Siegert shifts, in the control design is a complex task. Moreover, the coexistence of many motional modes poses additional complications.
Therefore, in the described scenario, \sota\ methods can become intractable.
This type of disturbance is typically mitigated during the calibration stage, where the uncertainties in infidelity \Inf\ are inferred and bounded.
 
These effects become more pronounced when the Raman laser intensities are increased to reduce the gate duration $\tau_\text{g}$.
Finally, \sota\ methods use gradient-based search methods to find feasible solutions while ignoring the highly non-convex landscape of the solution space.

\subsubsection{Control task for \twoq\ implementations}
Specific requirements exist for the geometric and dynamical phases to achieve high-quality $(q,r)$-\twoq\ infidelities.
The evolution operator is given by
\begin{equation}
    \mathds{1}^{\otimes2} \pm \text{i} \chi_{qr} {\rm X}_{q} {\rm X}_{r},
\end{equation}
where~$\mathds1$ and~${\rm X}$ are the identity and Pauli-X gates, respectively; normalisation is implied and not explicitly given in all our state and operator expressions. 
To obtain maximal entanglement between the electronic states of the pair~$(q,r)$, the accumulated geometric phase~$\chi_{qr}$ must be equal to~$\nicefrac{\pi}{4}$.

To ensure that there is no residual entanglement between the motional modes and electronic states, all dynamical phases in the phase space must return to their initial positions. 
This condition is also known as the `closure' condition, implying that every motional trajectory closes its displacement in the phase space~\cite{CDM+14}.

\subsubsection{Mathematical description for \twoq\ in open dynamics}

The \mmc\ Hamiltonian is closely related to Eq.~(\ref{eq:Hj})~\cite{CDM+14, DLF+16}.
For the parameters listed in Table~\ref{tab:mmc_hamiltonian},
\begin{table}[]
    \centering
    \begin{tabular}{c|c}
    \hline \hline
       Parameter  & Symbol \\ \hline
        $\Omega_{\jmath}$ & Rabi strength\\
        $\mu$   & Laser detuning\\
        $\eta$  & Lamb-Dicke parameter\\
        $a$ ($\da{a}$) & destruction (creation) operator\\
    \hline
    \end{tabular}
    \caption{Parameters in the \mmc\ Hamiltonian.}
    \label{tab:mmc_hamiltonian}
\end{table}
then the spin-dependent coupling is generated by a pair of Raman laser, blue and red detuned from the transition energy level, impinging each target ion which generate 
the interaction Hamiltonian
\begin{align}
\label{eq:MS_Hamiltonian}
H_\text{MMC}(t)
=- &\sum_{\jmath}^{\{q,r\}}\Omega_{\jmath} \cos\left(\mu t\right) \\
\times&\left[\text{Y}_{\jmath}+ \text{X}_{\jmath} \sum_{k=1}^{N} \eta_{\jmath k} \left(a_{k} \e^{-\ii \nu_{k} t}+a_{k}^{\dagger} \e^{\ii \nu_{k} t}\right) \right]\nonumber,
\end{align}
where the indices~$\jmath$ and~$k$ spans over the gate ions and the motional 
modes, respectively~\cite{LDM+03}.

We consider non-unitary processes such as heating and dephasing of vibrational modes
as well as both Rayleigh and Raman photon scatterings and laser intensity fluctuations ~\cite{TMK+00, OIB+07}. 
The Born-Markov approximation for these dynamics~\cite{GZ04}
yields the quantum master equation (QME)
\begin{equation}
\label{eq:alwayslabelequations!}
\text{i}\hbar\dot{\rho}=\left[H_\text{SM}(t),\rho\right] + \mathcal{L}[\rho] ,\; \mathcal{L}
[\rho]=\sum_l\mathcal{L}_l[\rho],
\end{equation}
for~$\rho$ the state of the ions, and the second term being the Liouvillian superoperator~\cite{GZ04,T06}.
We implement the Liouville jump operators~\cite{VPS23} obtained with reasonable rate values~\cite{WMI+98, HRB08} and the experimental data. 
We integrate the QME to time~$\tau_\text{g}$.

There are several ways to unravel \qme, such as the Monte Carlo wave function and quantum state diffusion~\cite{Car09, Brun00, MKS+19}.
We specifically use the quantum state diffusion (QSD) technique~\cite{GP92,Brun00}, which is a type of quantum trajectory theory (QTT).
The QSD method involves the Wiener process which drives the stochastic evolution of the state vector $\ket{\psi}$.
QSD also accommodates the interpretation of continuous monitoring of any system's observable, which is a necessary feature given that our control scheme implies position monitoring of a spectator ion.

\subsection{Position monitoring}\label{sec:backg:ion_position}
Here, we explain the basis of ion-position monitoring in a linear Paul trap.
We succinctly describe in the following order the physical model and the fundamental theory of 
position monitoring of a single trapped ion and its state-of-the-art.

\subsubsection{Theory of position monitoring}

The spectator ion is weakly transversely illuminated with a laser with respect to 
trap axis, coupled to its electric-dipole moment~\cite{WF+25, CDB+21, CAP+21_apl}.
The scattered light is collected with a spherical lens and reflected back from 
a mirror positioned at a distance such that the travel time of the reflected 
photon is shorter than the transition lifetime~$\Gamma$.
The mirror position is controlled and stabilised by a piezo-electric transducer (PZT)
such that the ion is at the node of the standing wave. 
The superposed electromagnetic fields are detected in a photo-multiplier tube (PMT) whose 
photocurrent was processed using a spectrum analyser.

The position-monitoring Hamiltonian is described as follows:
The key in the present scheme for position monitoring is the coupling of the 
electric field in the mirror mode $\bm{E}_\text{m}$ with the atomic dipole moment between the ground \ael2S1 and excited \ael2P1 states $\bm{d}_{\rm eg}$. 
Similarly, the corresponding electric field for the background modes is $\bm{E}_\text{b}$. 
For $\sigma^{+}_{\rm eg}={\sigma^{-}_{\rm eg}}^\dagger$ the rasing and lowering electronic operator for the spectator ion, then
\begin{align}
H_{\rm{s}}=-\frac{1}{2}&\Omega_{\rm s}\e^{k_\textbf{s}x_s}\sigma^{+}_{\rm eg} \nonumber \\
&-\bm{d}_{\rm eg}\cdot\left[\bm{E}^{+}_\text{b}(x)+\bm{E}^{+}_{\text{m}}(x)\right] \sigma^{-}_{\rm eg}+\text{hc},
\end{align}
is the atom-field interaction Hamiltonian~\cite{SRZ05}.

At the photodetector, the electromagnetic field reflected from the back mirror (signal) interferes with the elastic scattering field from the ion (local oscillator), which is shown as sidebands of the carrier transition exactly at frequencies $\bm{\nu}$. 
This scheme resembles a homodyne measurement~\cite{BRW+06, ERSB01, DHB+22, CDB+21} in which the signal is amplified and shown as sidebands at frequencies $\bm{\nu}$ whose amplitudes are modulated by the motion of the ion.

\subsubsection{State-of-the-art experiments}

Since the original experiments~\cite{ERSB01,SRZ05} the most recent works have demonstrated a sustained improvement~\cite{BHS+13, CAP+21_prl, CAP+21_apl} in spatial precision.
The first work of this kind consisted of a single $^{138}$Ba$^{+}$ confined in a linear Paul trap with a back mirror to investigate the fluorescence interference with its mirror image~\cite{ERSB01}. They continuously resolved the position of the ion down to a resolution of 1.7~nm, which was more precise than the spread of the ground-state wave packet $\Delta x_0=7$~nm.
Sub-Doppler temperatures were achieved using the same setup. They tracked the motion of the ion, whose signal was fed back to the trap's electrodes~\cite{SRZ05,BRW+06} reaching a temperature $30\%$ lower than the Doppler limit.
With the advent of more sensitive electronics and measurement techniques, a more refined control system incorporating quadrature measurement and phase-locking of the motion of the ion with the trap electrodes was implemented~\cite{BHS+13}. A record resolution of $0.3\Delta x_0$ ($\Delta x_0=6$nm) was achieved, limited only by the shot noise in the fluorescence signal.
More recently~\cite{CAP+21_prl}, a time-tagging detection system was implemented to reconstruct the average \ps\ trajectories to the single quanta of motion and demonstrated the real-time measurement of a single ion's position in an average time window of $T_0 =8$~\us.

A theoretical approach consisting of a complete solid-angle detector and a reference electric field shows how to saturate the  Heisenberg limit~\cite{TFN19}. 
In the optical setup, the reference field was significantly stronger than the dipolar scatterer. 
They also demonstrated that for an increasing NA of a spherical mirror, they can approach the Heisenberg limit faster than state-of-the-art methods with optomechanical resonators~\cite{YGL13}.
More recently, in a similar optical setup~\cite{WF+25}, the authors employed a hollow spherical mirror to control the back-action recoil by suppressing the amount of scattered light at large scattering angles.
This suppression leads to the Heisenberg limit for a certain combination of mirror reflectivity and NA.

Although the wavelength is typically two orders of magnitude greater than the  spread of the motional ground state~$\Delta x_0$, what really matters is the maximal 
slope of the created standing wave between the back mirror and the ion. 
Because the Rabi strength increases linearly with the electric field in a standing wave configuration, a displacement of 1~nm causes an increment of 2\% in the nominal Rabi strength~$\Omega_{\rm s}$.
In general, the photon detection rate~$\Gamma_{0}$ is determine by calibration of the system and includes imperfect optics reflectivity, numerical apertures and 
detector efficiency. 
Currently, this rate is maintained at a moderate order of $\sim 6\times 10^4~\rad$, typically with a detection contrast of 37\%~\cite{CAP+21_apl}.

\section{Approach}\label{sec:approach}
Now that we have completed our survey of the key background information and \sota\ method, we explain how to tailor our closed-loop control scheme to devise robust 2QGs.
We explain our closed-loop position-monitoring model of trapped-ion dynamics for the \twoq\ design, its underlying mathematical description, and the methods employed therein.

\subsection{Physical model}\label{sec:appr:model}
We now describe our model of the closed-loop control scheme for the \twoq\ design.
First, we characterise our control scheme in terms of the basic elements of control theory.
Second, we describe the physical components of our closed-loop control scheme.
Third, we elaborate on the simulated ion-trap environment for the \twoq\ design.

\subsubsection{Control elements}

We employ the definitions of the plant, controller, and policy as given in~\cite{VDP+23}.
The plant comprises (i) all the elements necessary to load, stabilise, and cool the ion-crystal; 
(ii) all devices necessary to generate, stabilise, reference, and redirect laser light towards the ion crystal;
(iii) all devices and parts necessary to generate a closed-loop communication channel.

The controller is composed of acousto-optic modulators or deflectors, whose corresponding RF power controls the amplitude and frequency of the outgoing Raman lasers. 
The input to the controller is the pair $\left(x_k,p_k \right)$ and the output is a real signal directed to the RF power sources driving the bichromatic lasers.

The policy is a map between the pair $\left(x_k, p_k \right)$ and the corresponding amplitudes of the bichromatic Raman lasers. The RL agent is responsible for tailoring the policy via interaction with the plant, and the map is stored in an M-layered neural network (NN).

\subsubsection{Model description}

A linear Paul trap confines~$N+1$ alkali-like ions, with labels
\begin{equation}
    \jmath \in[N+1]:=\{1,\ldots,N+1\},
\end{equation} 
where the last ion is the spectator.
The ions are prepared in an electronic (meta)stable stationary state~$\ket0$. The spectator ion is treated 
as described in~\S\ref{sec:back:control_2qg}. The remaining ions are used for information processing. Any pair of ions are used to implement the \twoq\ whose qubit states, $\ket{F, m_{\rm F}}$, are physically 
encoded in the hyperfine states of $^{171}$Yb$^+$, i.e. $\ket{00}$ and $\ket{10}$ with a frequency separation of~GHz~\cite{LBM+15}
\begin{equation}
\omega_\text{hfs} \sim 2\pi\times 12.643(2).
\end{equation}

\begin{figure}
\centering
\includegraphics[width=\linewidth]{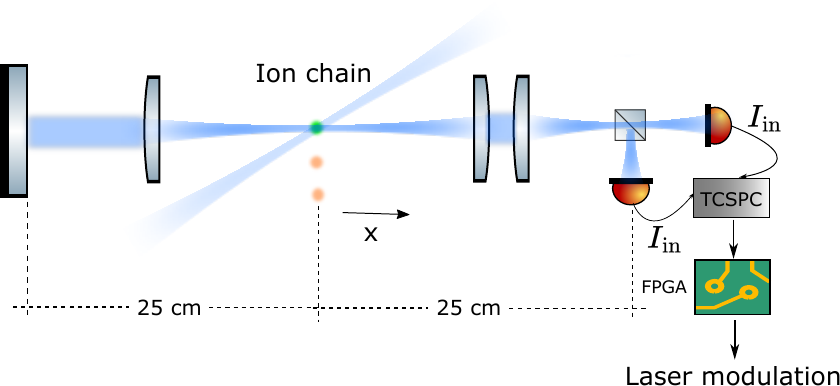}
\caption{Diagram of the CLPM-scheme. A three-ion chain is illustrated, where the orange dots represent gate ions and the green dot represents the spectator ion. The spectator ion is illuminated with an external coherent laser coupled to the \ael2S1 -- \ael2P1 transition, illustrated as a diagonal blue cone-like shape. 
An array of lenses and a back mirror collect and superpose the fluorescence light to undergo a two-channel time-correlated single-photon counting (TCSPC) setup. The output signal is sent to a FPGA for post-processing. 
The global and bichromatic Raman beams are not shown for simplicity.}
\label{fig:fb_model}
\end{figure}

The ion crystal is illuminated by a pair of Raman lasers blue an red detuned 
respect to any vibrational mode, whose shape is modulated~\cite{CDM+14}.~The stimulated Raman transfer is realized via the
\ael2P1 state. The global beam, is set up such that it illuminates the 
computational ions, and two pairs of blue and red Raman beams are tightly focused on the two gate ions. The position of the spectator ion 
is monitored when the \twoq\ gate implementation starts, and the control policy maps the reading of the position into blue- and red-detuned Raman laser intensities.
In Fig.~\ref{fig:fb_model}, we show the CLPM-scheme for a three-ion chain. The Raman lasers addressing the blue dots effecting the~\mmc\ gate are not shown for simplicity. 
The back mirror, separated by 25~cm from the spectator ion, reflects the fluorescence emitted by the ion such that it is superposed with the direct emission, and the photon travel time is shorter than the radiative lifetime of the dipole transition. 
The combined signal reaches a two-channel single-photon counting system which provides information about all the motional modes simultaneously ~\cite{CAP+21_apl}. 
Once the signal is generated, it is passed to an FPGA which contains the policy and generates the proper instruction to the optical devices responsible for changing the intensity of the bichromatic fields.

The coherent space of the motional modes is represented by the Fock space. Each motional mode has an autonomous assignment, the maximum value of which is seven.
The exact assigned dimension depends on the current value of the coherent state according to the actual dynamics, plus two additional phonons, capped at the maximum user-defined value $\bar{N}_{k}$.

\subsubsection{Simulated ion-trap environment}

We employ quantum trajectory theory (QTT) to unravel the \qme\ for the \twoq\ gate design. This method allows us to reduce the complexity from $N^2$ to $2N$ implying a significant reduction in time and memory to solve the QME. 
Another crucial reason is that
we can ascribe a subjective degree of reality to each quantum trajectory so that a learning agent can interact and learn about the dynamics of the plant. 
More specifically, we use the quantum state diffusion flavour of QTT to estimate the evolution of the wave-vector whose stochastic term is governed by a Wiener 
process~\cite{JS06, Brun00,GP92}.
We use the C++ `Quantum trajectory class library'~\cite{SB97} to integrate the resulting stochastic quantum trajectories.
We also developed our own code for the state-of-the-art (SotA) method for the \mmc~\cite{CDM+14,ZMD06}.
The simulated environment employs a time-adaptive Runge--Kutta--Cash--Karp $\nicefrac{5^{th}}{4^{th}}$ algorithm to solve stochastic differential equations.

Once the \ps\ trajectories and the infidelity are computed, they are passed on to the learning algorithm.
During the training stage, the algorithm evaluates the cost function which we discuss in~\S\ref{sec:appr:methods} and according to its internal logic, proceeds to update the neural networks (critic and actor), updates the intensity of the bichromatic beams, and waits for another episode.
During the production stage, the algorithm receives the \ps\ data points and uses the pre-stored NN to estimate the corresponding signal for the optical modulators, such that the intensity of the bichromatic fields is changed appropriately.

\subsection{Mathematical description}\label{sec:appr:math}
Here, we describe the mathematical interaction between the monitoring of the spectator ions and the computational ions during the \twoq\ implementations.
First, we describe the joint Hamiltonian that includes the \mmc\ and spectator ion Hamiltonians.
Second, we describe the noise and decoherence sources that are incorporated into our design.
Third, we explain how we account for the effect of monitoring a spectator ion.

\subsubsection{Hamiltonian}
The Hamiltonian~$H_\text{s}$ must be expressed in the interaction picture, similar to the \mmc\ Hamiltonian $H_\text{SM}$ in Eq.~(\ref{eq:MS_Hamiltonian}).
The transformation of the motional part into the interaction picture is standard and it is equivalent to a transformation from the \schr\ 
to Heisenberg picture. 
The transformation is complete by simply changing notation and bearing in mind that the new detuning $\Delta_\text{L}$ is referenced respect to~$\nu_{s}$
\begin{align}\label{eq:H_spectator}
H_\text{s} = &\left(\sigma^{+}_se^{-\ii\Delta_\text{L} t}+\sigma^{-}_se^{\ii\Delta_\text{L} t}\right)\\ \nonumber
&\times\left[\Omega_{\rm s}\e^{\ii(k_s \hat{x}_s)} - \bm{d}_{\text{eg}}\vdot\left( \bm{E}_\text{b}+\bm{E}_\text{m} \right) + \text{hc} \right]
\end{align}
which, after incorporating it in our full Hamiltonian we obtain
\begin{equation}
\label{eq:Monitoring_Hamiltonian}
H(t) = H_\text{MMC}(t) + H_\text{s}(t).
\end{equation}

As the motional modes evolve in an entangled state, directly applying the projective measurement~$\hat{x}$ on the wave-vector $\ket{\Psi}$
will lead to zero information, given that it depends on the electronic state of the gate ions. To extract the motional mode $\bar{x}_k:=\text{tr}(\dyad{++}{++} \otimes \hat{x}_k \rho(t))$ from the state vector, we need to employ the projection operator
\begin{equation}
    \Pi_{\imath\jmath k} =\sigma^\text{x}_{\imath}\sigma^\text{x}_{\jmath}\hat{x}_{k},
\end{equation}
where the gate ions are labelled as~$\imath$ and~$\jmath$. Similarly, for the other motional modes.

The maximum count rate is $\max{\Gamma_{0}} \sim 0.06~(\upmu \rm s)^{-1}$ for the position monitoring of a single trapped ion~\cite{CAP+21_prl}.
Such a rate provides approximately one photon every 
$\sim 16~\upmu$s for a total of eight \ps\ readings for a gate duration of 
$100~\upmu$s. In this work, the maximum count rate is increased to 1~$(\upmu \rm s)^{-1}$, which in turn means one event per \us. Based on experimental realisations the signal-to-noise ratio for the 
scheme proposed in~\cite{CAP+21_prl} is above 30~dB, implying sufficient room for a clean detection.

\subsubsection{Noise and decoherence}
Sources such as the AT and motional Bloch--Siegert shifts depend on the intensity of the bichromatic fields. 
The former is modelled by adding an offset to the effective frequency of the bichromatic field, whose value is $\alpha(\omega_{0,1})\Omega^2$, where $\alpha(\omega_{0,1})$ is the effective scalar contribution between the hyperfine levels $\ket{m_F=0, F=0}$ and $\ket{m_F=0, F=1}$. 
The latter contribution is already present in Hamiltonian Eq.~(\ref{eq:MS_Hamiltonian}). 
Similarly, the effect of the bare electronic transition of the gate ions increases with $\Omega_{\jmath}$ and is given by Eq.~(\ref{eq:MS_Hamiltonian}).

Decoherence sources are included as jump operators in the QME. These sources have been studied and included in our previous work~\cite{VPS23}.

Several noise sources can be directly translated into fluctuations in the motional frequencies.
Among them, the most relevant are the RF and end-cap trap voltage fluctuations.
We model this effect by adding an offset $\delta f$ to the motional frequencies described by a Gaussian distribution with a mean of zero and variance $\sigma_f$.

\subsubsection{Impact of position-monitoring of spectator ion}

We use the truncated Dyson series, given $\mathcal{T}$ as the time ordering operator
\begin{align}\label{eq:truncated_dyson}
    U = \mathds{1} - \ii &\int_0^t \dd t_1 H(t_1) \nonumber \\
    &- \frac{1}{2}\mathcal{T}\int_0^t\int_0^t \dd t_1 \dd t_2 H(t_1)H(t_2)+O(t^3),
\end{align}
where $O$ is the round-off error due to higher-order terms in the series, which determines the evolution of Eq.~(\ref{eq:Monitoring_Hamiltonian}).
The resulting terms of the series allow us to quantify the various effects owing to the continuous position monitoring of the spectator ion on the \Inf.

To understand the different interactions that affect \twoq\ due to position monitoring, we effect the multiplication term by term in Eq.~(\ref{eq:MS_Hamiltonian}) and Eq.~(\ref{eq:H_spectator}) in the truncated Dyson expression, Eq.~(\ref{eq:truncated_dyson}).
This led us to identify three types of interactions: (i) the cross terms between gate ions' and spectator's bare electronic transitions, (ii) the product between the terms involving the motional modes excitation from the Raman lasers and the external driving laser, and (iii) the cross term between the mirror modes of the spectator ion and the second term in Eq.~(\ref{eq:MS_Hamiltonian}).

The jump operator for the spectator ion $s$ and $N$ motional modes have the basic form~\cite{SRZ05}
\begin{equation}
    L_{\rm PM}^{(k)} = \sqrt{\frac{\Gamma_{\rm m}}{2}}\eta_{s k} ( a_k + a_k^\dagger).
\end{equation}

\subsection{Methods}\label{sec:appr:methods}
We now describe the methods employed to solve the control problem.
First, we cast the \twoq\ design as a learning problem.
Second, we describe the reinforcement learning algorithm employed in our proposed control scheme.
Third, we explain how we construct the cost function.

\subsubsection{Casting \twoq\ design as learning}
We introduce continuous monitoring of the spectator ion in a chain of~$N+1$ ions to reconstruct~$\alpha_{\jmath k}(t)$ in real-time.
This information is used to devise robust control policies for implementing \twoq\ against motional heating, drift in motional frequencies, and scattering processes.
We also discretise time by introducing a time mesh comprising~$m$ equal segments so $t\in(\tau/m)\mathbb{Z}_{m+1}$
with~$m$ a hyperparameter, that is, a parameter that is `tuned' for optimisation.
As $m= 2N+1$ suffices~\cite{CDM+14,ZMD06}, we replace~$\Omega(t)$ with ~$\bm\Omega\in[0,\Omega_\text{max}]^m$.

To cast a control problem into a RL problem, we identify the components of a control task as the components of RL~\cite{VDP+23}, namely, environment~E, set of environment states~$\bm S$, agent~A, set of agent's actions~$\bm A$ and reward~$R$.
For our \twoq\ RL problem, we define E as the ion trap device, which holds a linear chain of~$N+1$ trapped ions and all the equipment required for its operation.
The states of~E are defined by the set~$\bm S = \{(x_{kt_i},p_{kt_i},\bm\nu,\mu,\bm \Gamma)\mid t =\tau_{\rm g}\}$.
~A is an algorithm that runs on the processing unit that controls~$\Omega(t)$ and its actions are defined as $\bm A = \{ a_t \in [0,\Omega_\text{max}] \mid t \in ?\}$.
We define the immediate reward after transition from~$s_t$ to~$s_{t+1}$ with action~$a_t$ as the estimate of the collected geometric phase 
at~$s_{t+1}$,~$R_{a_{t}}(s_t,s_{t+1}) := \tilde{\chi}_{qrs_{t+1}}$. We define the episodic reward, which is given to the agent at the end of \twoq\ 
implementation, that is, ~$t=\tau_{\rm g}$, as~$R_{\tau_{\rm g}}=\mathcal{I}$.

A basic RL agent interacts with its environment at discrete time intervals. At each time~$t$,
the agent receives the current state~$s_t$ and reward~$r_t$. Then, an action~$a_t$ is selected from the set of available actions and sent to the environment. The environment moves to a new state~$s_{t+1}$ and the reward~$r_{t+1}$ associated with the transition~$(s_t, a_t, s_{t+1})$ is determined. We now formally present the \twoq\ design problem as a RL problem.
\begin{task}[\twoq\ control as RL]
\label{task:2qg_learning}
Input:
\begin{itemize}
    \item[] $N+1$ trapped ions,
    \item[] gate ions labelled~$(q,r)$ and spectator ion $s$ whose electronic states are initialized in~$\ket0$,
    \item[] gate duration $\tau_\text{g}$, 
    \item[] $N+1$ motional angular frequencies~$\bm\nu$,
    \item[] physical bounds for~$\Omega_{q}(t) = \Omega_{r}(t) = \Omega(t)$, 
    \item[] maximum detuning drift $\delta_\text{tol}$, 
    \item[] target infidelity $\mathcal{I}_\circledcirc$, 
    \item[] signal latency $T_0$, and
    \item[]and the fluorescence emission rate of the spectator ion $\Gamma_{0}$,
\end{itemize}
Output: Devise a policy $\pi$ delivering pulse sequences $\Omega(t)$ such that $\mathcal{I}$ of the final state for pair of ions $(q,r)$ is within some given distance $\epsilon$ from the Bell states $\Phi^{\pm}$.
\end{task}

With the task described above the problem to solve becomes,
\begin{problem}[Feasibility]
\label{prob:phys}
Find solutions to task~\ref{task:2qg_learning} which yield output:  $\Omega(t)$ and emission rate of the spectator ion $\Gamma_\text{m}$ s.t.
\begin{equation}\label{eq:feas1}
    0\leq\Omega_\jmath\leq\Omega_{\rm{}{max}}~\forall\jmath,
\end{equation}
and scattering rate
\begin{equation}\label{eq:feas2}
    \Gamma_{\rm m}\geq \Gamma_{0},
\end{equation}
signal latency $T_{\rm CLPM}$  satisfying
\begin{equation}\label{eq:feas3}
    T_{\rm CLPM} \leq T_{0},
\end{equation}
and
\begin{equation}
\label{eq:feas4}
\underset{\left|\delta\right| \le \delta_\text{tol}}{\rm{max}}~\mathcal{I}(\delta) \le \mathcal{I}_\circledcirc,\,
\mathcal{I}:=1-\bra{\Phi^{\pm}}\rho(\bm\Omega,\mu,\delta,\tau)\ket{\Phi^{\pm}},
\end{equation}
\end{problem}

\subsubsection{Choice of reinforcement learning heuristic}

The motivation is to have an algorithm with data efficiency and reliable performance of a trusted region policy optimisation algorithm (TRPO)~\cite{LH24}, while using only a first-order expansion.
We pass~$(x,p)$ to a RL algorithm called Proximal Policy Optimisation (PPO)~\cite{SWD+17} to confection policies. 
The  search for a feasible~$\bm\Omega$ and~$\mu$ in the region $\left[0, \Omega_{\rm max} \right]^m \times \left[\mu_\text{min}, \mu_\text{max} \right]$ over the uniform measure.

Although there is no recipe for designing cost functions, each method here obeys certain heuristics and intends to balance the number of actions and the minimum required number of pulses according to the \sota\ equal to $2N+1$.  
We approach the training based on the actual observation of the \ps\ trajectories and the hypothetical actual geometric phase.
The gist of this approach lies in providing the learning agent with a sense of the intensity of the laser being used such that it gets as close as possible to $\nicefrac{\pi}{4}$ at the end of the gate duration. 
Finally, the agent is rewarded based on how low \Inf\ is at the end which assesses the closure condition.

If we used as a baseline $\Delta t_i = 8~\upmu$s as discussed above, that will provide us 
with approximately ten actions to be executed during a \twoq\ implementation of $100~\upmu$s. For higher rates $\Gamma_{0}$ more actions could be 
employed a per-gate implementation. This training also involves the index of the observation which encourages the agent to use more laser intensity at the early stages of the gate but the contrary when the 
end approaches.

\subsubsection{Cost function}\label{sec:appr:met:cost}
We approach the training from a different angle based on the actual observation of the \ps\ trajectories and the hypothetically 
actual geometric phase. The gist of this approach lies in providing to the learning agent a sense of the intensity of the laser being used is such that 
it gets as close as possible to $\nicefrac{\pi}{4}$ at the end of gate time. Finally, the agent is rewarded based on \Inf.

The cost function involves the index of the observation which encourages the agent to use a higher laser intensity at the early stages of the gate; however, the opposite occurs when the end approaches.
The RL agent receives $(x_i,p_i)$ to determine the radius  where the cost function is
\begin{equation}
R := \pi\abs{\sum_k a_k\left(x_{kt_i}^2 + p_{kt_i}^2\right)}-\frac{\nicefrac{\pi}{4}}{\text{i}}.
\end{equation}
Notice that the geometric phase does not accumulate over the course of the gate time, on the contrary, the geometric phase condition is evaluated on the actual reading 
where index~$i$ indicates the early and final readings of the \ps\ trajectories.

\section{Results}\label{sec:results}
We present our results in three subsections. The first concerns the feasibility of the proposed control scheme. We then present the effects of spectator-ion position monitoring on \Inf. Finally, we demonstrate the performance of a reinforcement-learning algorithm for \twoq\ implementations.

\subsection{Feasibility of the closed-loop control} \label{sec:res:feasibility}
Here,  we demonstrate the feasibility of our CLPM-scheme.
First, we show that the signal rate $\Gamma_\text{m}$ meets the feasibility condition in Eq.~(\ref{eq:feas2}).
Second, we show that the signal latency and processing time $T_{\rm CLPM}$ also satisfy the condition in Eq.~(\ref{eq:feas3}).
Finally, we show that the \ps\ trajectories can be reconstructed under the current signal acquisition time and laser-power constraints.

\subsubsection{Signal rate}\label{sec:res:feas:strength}
The maximal slope of the standing wave between the back mirror and the spectator ion plays a crucial role in position-monitoring schemes. 
The Rabi strength increases linearly with the electric field in a standing wave configuration; hence, a displacement of 1~nm causes an increment of 2\% in the nominal Rabi strength~$\Omega$. 
Assuming that the dipole transition is not 
saturated, a typical displacement of $\Delta x = 2\eta \Delta x_0 \Omega/\Delta_\text{L}\approx 50\Delta \text{x}_0$ will be perceived as an increased Rabi rate of $50\Delta x_0 k = 50\eta$, which for a typical $\eta=0.05$ we obtain $\approx 2.5 $ times of an increment in 
the detected photons.

\begin{figure}
\centering
\begin{tabular}{cc}
\includegraphics[width=\columnwidth]{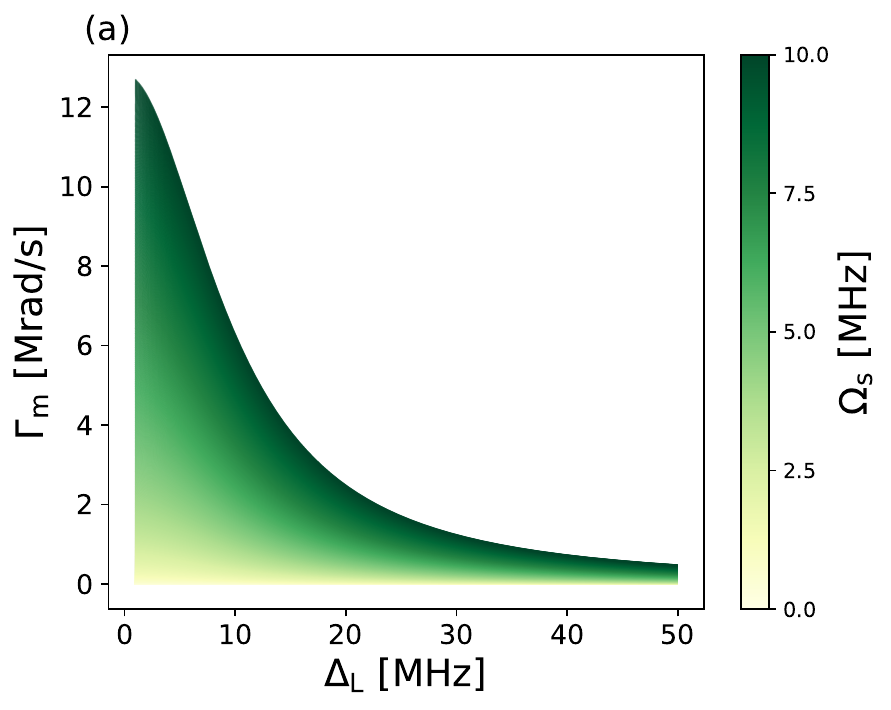}\\ \includegraphics[width=\columnwidth]{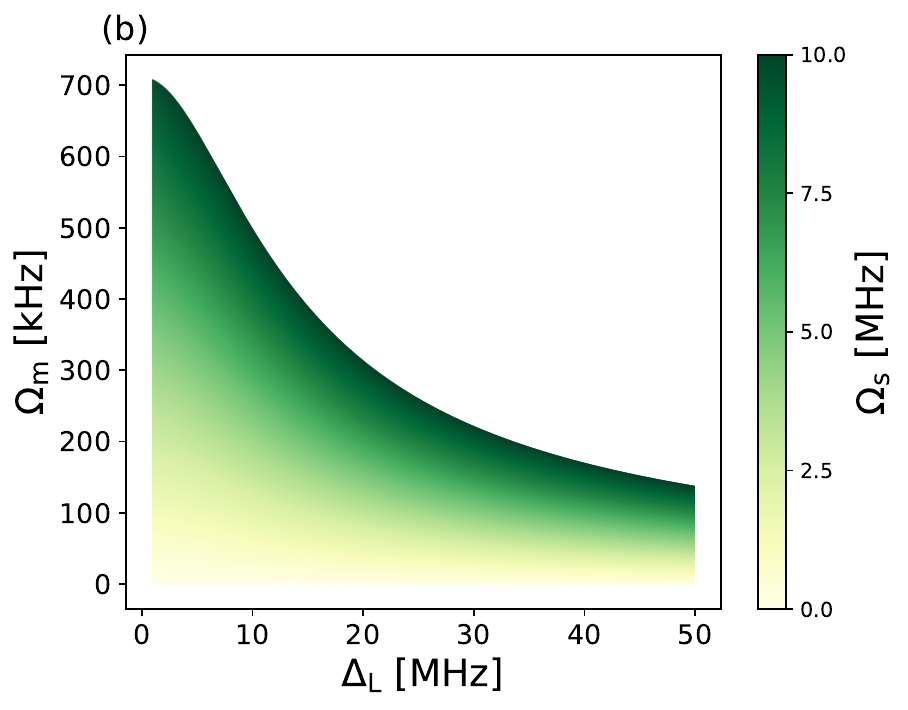}
\end{tabular}
\caption{(a) Emission rates of the spectator $\Gamma_\text{m}$ ion, and (b) Rabi strength in the mirror mode $\Omega_\text{m}$ versus external laser detuning $\Delta_\text{L}$. The colour map reflects the Rabi strength of the spectator ion $\Omega_s$.
 The parameters used are ~$\text{NA}=0.4$, $w_0=2~\upmu$m, $\Omega_s=3$~MHz.}
\label{fig:rabi_rates_mirror}
\end{figure}

The back-reflected electromagnetic field is coupled to the electric dipole moment of the spectator ion. 
The coupling strength depends on the numerical aperture (NA) of the collecting lens and the mirror reflectivity.
In Fig.~\ref{fig:rabi_rates_mirror} we show the Rabi strength in the mirror modes exerted on the spectator ion.
The colour map relates to the Rabi pulse shape acting on the spectator ion $\Omega_s$. 
Fig.~\ref{fig:rabi_rates_mirror}(a) shows the back-reflected fluorescence light collected by the lens. 
Fig.~\ref{fig:rabi_rates_mirror}(b) shows the effective Rabi strength of the mirror modes coupled to the spectator ions. 
The colour map on both plots relates to the initial Rabi strength $\Omega_s$ of the external laser driving the spectator ion.

In general, the photon detection rate~$\Gamma_{0}$ is measured by calibration of the system and includes imperfect optics reflectivity, numerical apertures and 
detector efficiency.
Currently, this number is kept in a moderate order of $\Gamma_0 \sim 6\times 10^4~\rad$, typically with a detection contrast of 37\%~\cite{CAP+21_prl}; nonetheless, we push this limit to $\Gamma_{\rm CLPM} = 1\times 10^5~\rad$.
In this work we do not consider detection efficiencies and optical imperfections, our core interest is to prove that our scheme can effectively compensate for typical errors and decoherence sources in real time.

\subsubsection{Signal latency and processing time}\label{sec:res:feas:latency}
The total path travelled by fluorescence photons 
until they reach the photodetector, as illustrated in Fig.~(\ref{fig:fb_model}). 
The photon travel time until it reaches the spectator ion is 1.6~ns and 2.4~ns more to the photodetector.
From the information above and given a lifetime of $\sim 8.7$~ns, we can safely conclude that the inherent latency of the closed-loop control is much lower than the gate duration, which is typically a few hundred microseconds.

In the limit where the feedback signal delay goes to zero, Markovianity of the plant is 
preserved~\cite{DZ02}. Wiseman~\cite{Wis94} showed that a response time much lower than the fundamental process to be controlled,
which in our scheme, it is the quantum logical gate duration, which also preserves Markovianity.
The typical \twoq\ gate duration is linked to several factors, such as the Lamb-Dicke parameter, the number of ions, and more importantly the laser 
detuning of the bichromatic Raman beams. Typically, the detuning is on the order of $\mu \sim 10$~kHz giving a total gate duration of $100~\upmu$s 
to execute a full turn such that the \ps\ trajectories are closed. This period is much longer than the response time of the feedback signal, as shown in the previous paragraph.

The computational complexity of neural network forward passes is commonly expressed in terms of the number of floating-point operations (FLOPs).
For fully connected layers, each weight typically contributes with a single multiply--add 
operation, yielding approximately $2 \times m \times n$ FLOPs for an input dimension $m$ and an output dimension $n$. Applying this calculation to our network architecture with an input dimension of six, hidden layer sizes 
of 256 and 512 for the Actor and Critic, respectively, and an output dimension of nine, resulting in approximately 12.6 kFLOPs per forward pass for the actor and 7.7 kFLOPs for the critic. Although this theoretical computational 
cost remains consistent across hardware platforms, the practical performance can vary significantly due to differences in parallelization, pipeline efficiency, and memory bandwidth, particularly when comparing general-
purpose processors with FPGA-based implementations.

\subsubsection{Reconstruction of phase-space trajectories}\label{sec:res:feas:phase_space}
In our control scheme, the position and momentum  of the motional modes are extracted from the wave-vector state resolved using quantum trajectories. 
The computation is performed continuously, mimicking continuous monitoring, although the actions are taken at intervals $\Delta t_i$. 
Because the motional modes evolve in a superposition state, directly applying the operator $\hat{x}$ will lead to zero, owing to its dependence on the electronic state of the gate ions.
However, the correct approach is via joint operators, that is, composite operators with internal and external degrees of freedom.

To determine the motional mode $\hat{x}_{\text{CoM}}$, it would be required the following projection operator, $\Pi_{\text{CoM}} =\sigma^{\rm x}_{\imath}\sigma^{\rm x}_{\jmath}\hat{x}_{\text{CoM}}$,
where the gate ions are labelled as $\imath$ and $\jmath$. We proceed similarly for momentum operators.
As only the position is measured through fluorescence detection, the period of time $\Delta t_i$ can provide sufficient information to the experimentalist to filter the data and numerically compute the momentum~\cite{ACC+21}.

To provide position and momentum readings to the learning agent, we take $8$~\us\ as a baseline, which translates into $\sim 80$~MFLOP to post-process and generate an action by the learning agent.
In Table~\ref{tab:time_scales} we show the different time scales associated with the ion trap dynamics.

\begin{table}
\centering
\caption{Relevant time scales for ion-trap operations.}
\begin{tabular}{c|c}
\hline
\hline
Parameter & value\\
\hline
Trap frequency $\left[\nu\right]$ & $\sim~3$~MHz ($\leq$1 \us)\\
Detuning $\left[\mu\right]$ & $\sim~10$~kHz (100 \us\ period)\\
FPGA $\left[\rm FLOP \right]$ & $\sim 10$~MFLOP (per \us)\\
Agent actions  $\left[\Delta t_i \right]$   & $\leq 10$~\us\ \\
\hline
\end{tabular}
\label{tab:time_scales}
\end{table}

\subsection{Impact of spectator ion's monitoring on $\mathcal{I}$} 
\label{sec:res:impact}
Now, we quantify the impact introduced by the monitoring of the spectator ion on $\mathcal{I}$.
We begin by demonstrating the impact of the spin-spin coupling between the gate ions and the spectator ion.
Second, we show the effect of motion-mediated errors due to spectator ion monitoring.
Third, we demonstrate the impact of the mirror-mode electromagnetic field coupled to the dipole moment of the spectator ion.

\subsubsection{Impact of spin coupling}\label{sec:res:imp:spin}
In this work, we demonstrate the validity of our control scheme in a three-ion chain confined in a linear Paul trap, whose transverse motional frequencies along the x--axis are
\begin{equation}
\label{eq:motionalfreqs}
\bm{\nu}_x \in \{4.3807, 4.3414, 4.2857 \}~\text{MHz}.
\end{equation}
The gate is mediated via transverse motional modes, which offer tighter confinement, thereby reducing the gate duration.

After expanding in Dyson series and identifying the spin interaction in the \twoq\ evolution (see Appendix~\ref{appendix:A}), the net effect on $\mathcal{I}$ is given by
\begin{equation}\label{eq:error_spin}
\mathcal{I}_\text{spin}=\sum_{\jmath}^{q,r}\frac{\Omega_{\rm s}\Omega_{\jmath}^{\rm max}}{\mu\Delta_\text{L}}\kappa_\text{c},
\end{equation}
where $\kappa_{\text{c}} = \int^\tau_0 \dd t \Omega_{\jmath}\cos(\mu t)$, which is very similar in form to the spin interaction between the two gate ions \cite{MC+21}. 
The factor $\kappa_\text{c}$ accounts for the error due to the closure condition of the \ps\ trajectories, which is a consequence of the bare electronic transition modulation or carrier effect of the gate ions \cite{SM99, SM00, BML24}.

\begin{figure}[h!]
\centering
\includegraphics[width=\columnwidth]{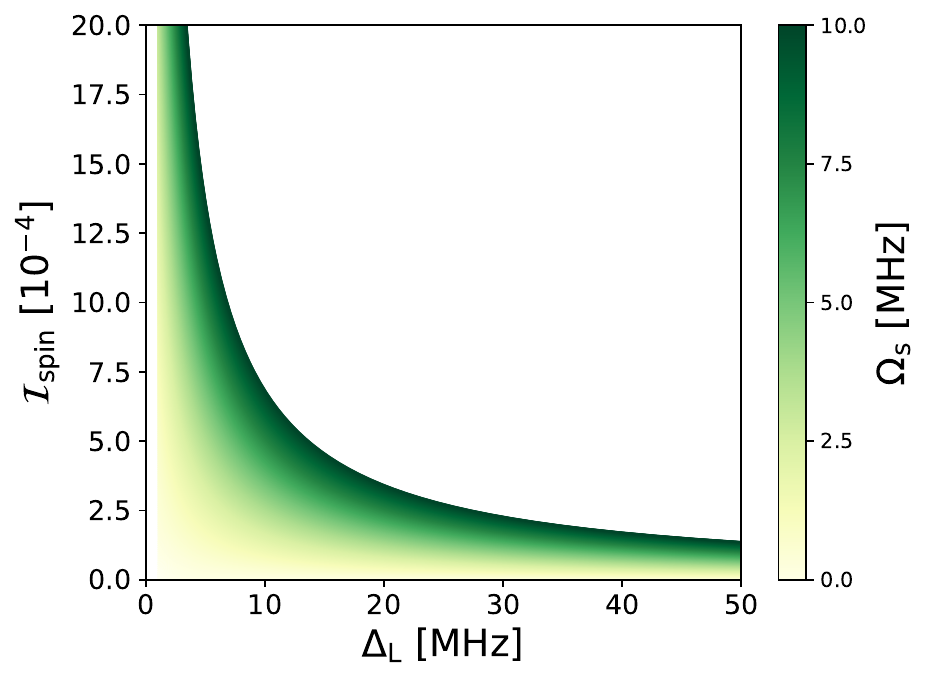}
\caption{Error in \twoq\ implementation due to spin coupling in a chain of three ions. $\Delta_\text{L}$: laser detuning of the spectator ion. $\Omega_{\rm s}$: Rabi strength on the spectator}
\label{fig:spin_error}
\end{figure}

In Fig.~\ref{fig:spin_error} we show $\mathcal{I}_\text{spin}$ where we have 
considered the conservative factor $\kappa_\text{c} \sim 1\%$ to account for errors due to bare electronic modulation when designing the pulse shape.
Here, we observe the decay of $\mathcal{I}_{\rm spin}$ with the inverse of $\Delta_\text{L}$.

\subsubsection{Motion-mediated entanglement}\label{sec:res:imp:motion}

The motional modes are also excited via dipole-transition driving of the spectator ion. 
The type of excitation that takes place here involves a different transition from the Raman lasers on the gate ions, namely, the \ael2S1--\ael2P1 dipole transition. 
The excitation of the collective motional modes via the spectator ion disturbs the entangling operation of the \mmc\ protocol.

\begin{figure}
\centering
\includegraphics[width=\columnwidth]{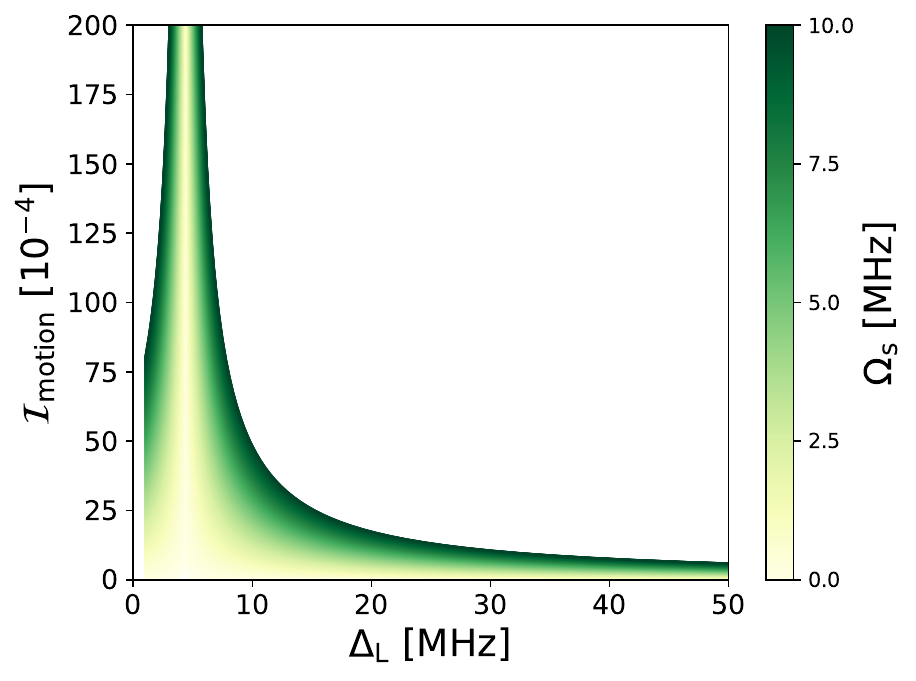}
\caption{Error in two-qubit gate due to motion-driven entanglement in a chain of three ions. $\Delta_\text{L}$: laser detuning of the spectator ion. $\Omega_{\rm s}$: Rabi strength on the 
spectator. The peak coincides with the motional mode frequencies (not including the remaining two modes for clarity).}
\label{fig:motion_error}
\end{figure}

The residual entanglement between the electronic states of the gate ions and the motional modes is given by
\begin{equation}\label{eq:error_motion}
\mathcal{I}_\text{motion} = \sum_{kk'}\sum_{\jmath}^{q,r} \frac{\eta'_{sk'}\eta_{\jmath k} \Omega_{\rm s}\Omega_{\jmath}^{\rm max} \alpha_{\tau}}{\abs{\left(\Delta_\text{L}-\nu_{k'} \right)\left(\mu - \nu_k  \right)}},
\end{equation}
where we have incorporated a damping factor $\alpha_\tau$ owing to the closure condition of the \ps\ trajectories. This expression evidences the trade-off 
between the emission rate $\Gamma_\text{m}$ and the quality of the \twoq\ implementation.
In Fig.~\ref{fig:motion_error} we show the infidelity as a function of the laser detuning $\Delta_\text{L}$.

\subsubsection{Impact of mirror-mode coupling}\label{sec:res:imp:mirror_mode}

The motion of the spectator ion gets coupled to the mirror mode via its dipole moment $d_{\rm eg}$ and the quantum noise 
operator~$b_\text{m}$. In the regime where the photon flight-time is much smaller than the natural decay rate of the dipole 
transition \ael1S1--\ael2P1, we can consider the closed-loop control as a Markovian process and conveniently drop out any time delay from now on.

The residual spin entanglement from the interaction between the mirror-reflected electromagnetic field and the electric dipole moment of the spectator ion involves Wiener processes. This stochasticity is also transferred to the motional modes via recoil, introducing new nuances in the \twoq\ dynamics.
In Appendix~\ref{appendix:A}, we detail the calculations, where we use some approximations to bound the error, such as the Lamb-Dicke regime and the commutation between the noise operators and the system operators.

With all those considerations above, given the gate duration $\tau_\text{g}$, the intensity of the laser coupling on the spectator in 
$\Omega_s$, the Lamb-Dicke parameters of the gate ion $\jmath$ and the spectator ion $s$ and the mean occupation number of the $k^{\rm th}$ motional mode $\bar{n}_k$ we obtain a maximum error given by
\begin{equation}\label{eq:error_mirror}
\mathcal{I}_\text{mirror} =\sum_{kk'} \sum_{\jmath}\frac{\eta'_{sk'}\eta_{\jmath k}\Omega^3_\text{m} \Omega_{\jmath}^{\rm max} \alpha_{\tau}}{\abs{\mu-\nu_k}\sqrt{\left(\Delta_\text{L}-\nu_{k'}\right)^5}},
\end{equation}
where we can observe the interplay between all the elements in the experimental layout and the atomic element properties.

\begin{figure}
\centering
\includegraphics[width=\columnwidth]{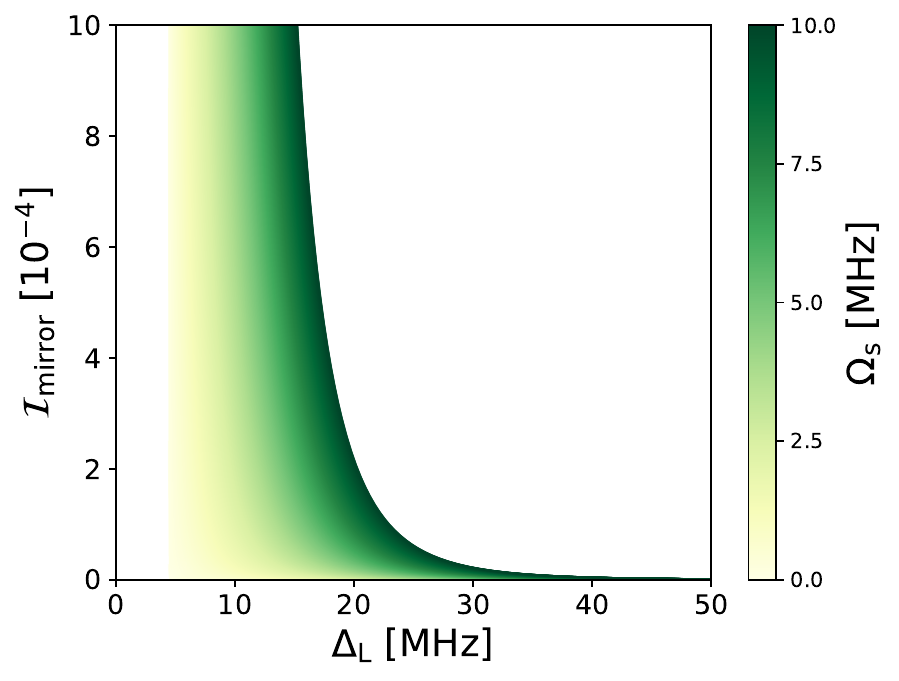}
\caption{Error due to the mirror modes coupled back to the spectator. The parameters are the same as in Fig.~\ref{fig:motion_error}.}
\label{fig:mirror_modes}
\end{figure}

In Fig.~\ref{fig:mirror_modes} we show the total error due to the electromagnetic mirror mode.

In Table~\ref{tab:infidelity_cont_mon} we show the combined fractional errors due to position monitoring with respect to other sources of error. We compare the corresponding fraction of the errors $\mathcal{I}_\text{spin}$,
$\mathcal{I}_\text{motion}$, and $\mathcal{I}_\text{mirror}$.
With respect to motional drift and heating in isolated cases, the former delivers \Inf$=(150\pm5)\times10^{-4}$, while the later delivers \Inf$=(92\pm3)\times10^{-4}$.

\begin{table}[H]
\centering
\caption{Error due to continuous monitoring at $\Omega_{\rm s}=10.0$~MHz, $\Delta_\text{L} = 32$~MHz. For $\text{NA}=0.4$, $w_0=2~\upmu$m, $\Gamma_{\text{MH}} = 100$~1/s and  $\Delta\nu_k = \pm 600$~Hz, $\bm{\nu}_x = \{4.3807, 4.3414, 4.2857 \}$~MHz, $\alpha_{\tau}=0.01$.}
\begin{tabular}{c|c|cc}
\hline  \hline
\multirow{2}{*}{Type (Symbol)} & \multirow{2}{*}{Error [$\times 10^{-4}$]}  & \multicolumn{2}{c}{Fractional error}\\
 & & Heating [\%]  & Drift [\%] \\
\hline
$\mathcal{I}_\text{spin}$ & 1.5 & 1.0 & 1.6 \\
$\mathcal{I}_\text{motion}$ & 12.0 & 8.0 & 13.0 \\
$\mathcal{I}_\text{mirror}$& 0.4 & 0.3 & 0.4 \\
\hline
\end{tabular}
\label{tab:infidelity_cont_mon}
\end{table}

\subsection{Data-driven \twoq\ control} \label{sec:res:learning}
Here, we explain the results of the \twoq\ implementations, following our closed-loop position-monitoring control scheme. First, we describe how the RL agent is trained.
Second, we show the results of the RL agent trained by following episodic rewards based on $\mathcal{I}$ at~$\tau_\text{g}$.
Third, we show the results of the RL agent trained by following a geometric-phase-based cost function and episodic reward and~$\mathcal{I}$ at~$\tau_\text{g}$.

\subsubsection{Reinforcement-learning agent training}\label{sec:res:lear:training}

We train the RL agent by running quantum trajectories with the corresponding jump operator for position monitoring at~$\Gamma_\text{CLPM} = 2\times10^5~\rad$. 
Additionally, we incorporate the carrier term, AT and BS energy shifts, and motional heating at rates $\Gamma_{\rm H}=100~\rad$ for all motional modes. 

In Table~\ref{tab:hyperparameters}, we list the set of hyperparameters that were customised for this study.
In all the training scenarios presented here, we `encourage' the learning agent to explore `more' by increasing the `Entropy' parameter while reducing the learning rate for both the critic and actor networks.
In this sense, a larger Epoch and Batch were required coherently with the size of the neural networks to ensure a more robust learning, whereas from experience, the clipping parameter was determined to observe convergence in the training stage.

Using the set of parameters provided in Table~\ref{tab:hyperparameters}, we observed convergence after approximately 12,000 episodes. The duration of each episode strongly depends on the choice of Lindbladian operators used during the training or production stage and the number of ions involved.
In the case presented here, the computer simulation of each episode takes an average of 1~s, including both the decision-making process of the learning algorithm and the corresponding neural-network update.
During training, after convergence is reached, the performance typically remains approximately stationary for a further 2,000 episodes before degrading sharply. When this degradation entails the loss of previously acquired behaviour as a consequence of continued or sequential learning, it is consistent with catastrophic forgetting, namely, interference with knowledge acquired during the earlier stages of training~\cite{McCloskeyCohen1989,French1999}. Such forgetting can be mitigated using established continual-learning techniques, including rehearsal or replay methods and parameter-regularisation strategies~\cite{Robins1995,Atkinson2021,Kirkpatrick2017,Parisi2019}. The reported time measurements were specific to the hardware configuration used in the simulations and could vary across the computational setups.

\begin{table}
\caption{Nominal hyperparameters values for the training of PPO}
\vspace{10pt}
\begin{tabular}{c|c}
\hline
\hline
    Nominal value & Description\\
    \hline
    Learning rate for critic models     & 0.0004  \\
    Learning rate for critic models     & 0.0001  \\
    Clipping surrogate objective        & 0.185     \\
    Rate entropy into the loss function & 0.007   \\
    Discount factor                     & 0.95     \\
    Generalized advantage estimation    & 0.8     \\
    Number of update                    & 128     \\
    Batch size for sampling             & 128     \\
    Number of rollout                   & 500     \\
    Neural network depth for actor      & 256     \\
    Neural network depth for critic     & 512     \\   
    \hline
\end{tabular}
\label{tab:hyperparameters}
\end{table}

\subsubsection{\twoq\ employing episodic rewards}\label{sec:res:lear:episodic}

An episodic approach is always interesting to observe in any reinforcement learning application. 
In our case, we let the algorithm to only observe the \ps\ trajectories and execute actions but with no reward until the end of the gate duration $\tau_{\rm g}$. 
We are interested in assessing whether the algorithm can find a different dynamic and excursion of the motional modes, such that the task is accomplished by minimising the  disturbances intrinsic to the open ion-trap dynamics.

\begin{figure}
\centering
\includegraphics[width=0.6\linewidth]{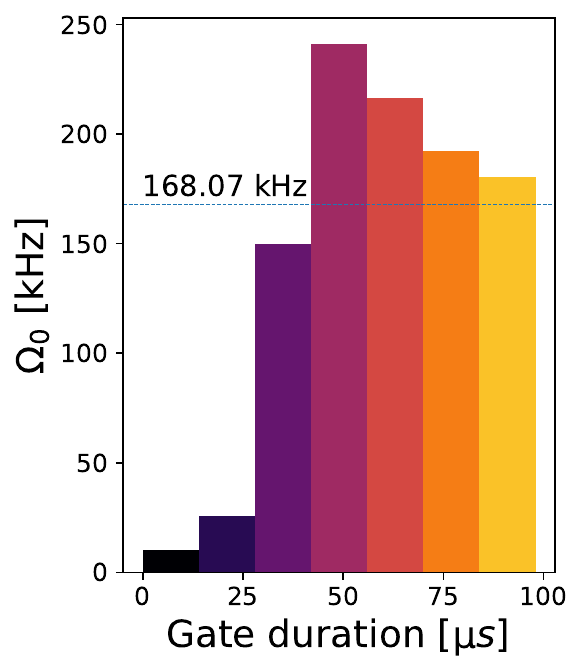}
\caption{Average seven-segments piece-wise constant pulse shape devised employing episodic rewards. The colours are arbitrary and intended to guide the eye over each piecewise constant segment. 
The blue dashed horizontal line represents the RMS value of the pulse sequence.}
\label{fig:pulse_episodic}
\end{figure}

In Fig.~\ref{fig:pulse_episodic} we show a seven-segment piece-wise constant pulse shape. 
The maximum Rabi strength is~$238$~kHz and the RMS value is~$168.07$~kHz.

\begin{figure}
\centering
        \includegraphics[width=1 \linewidth]{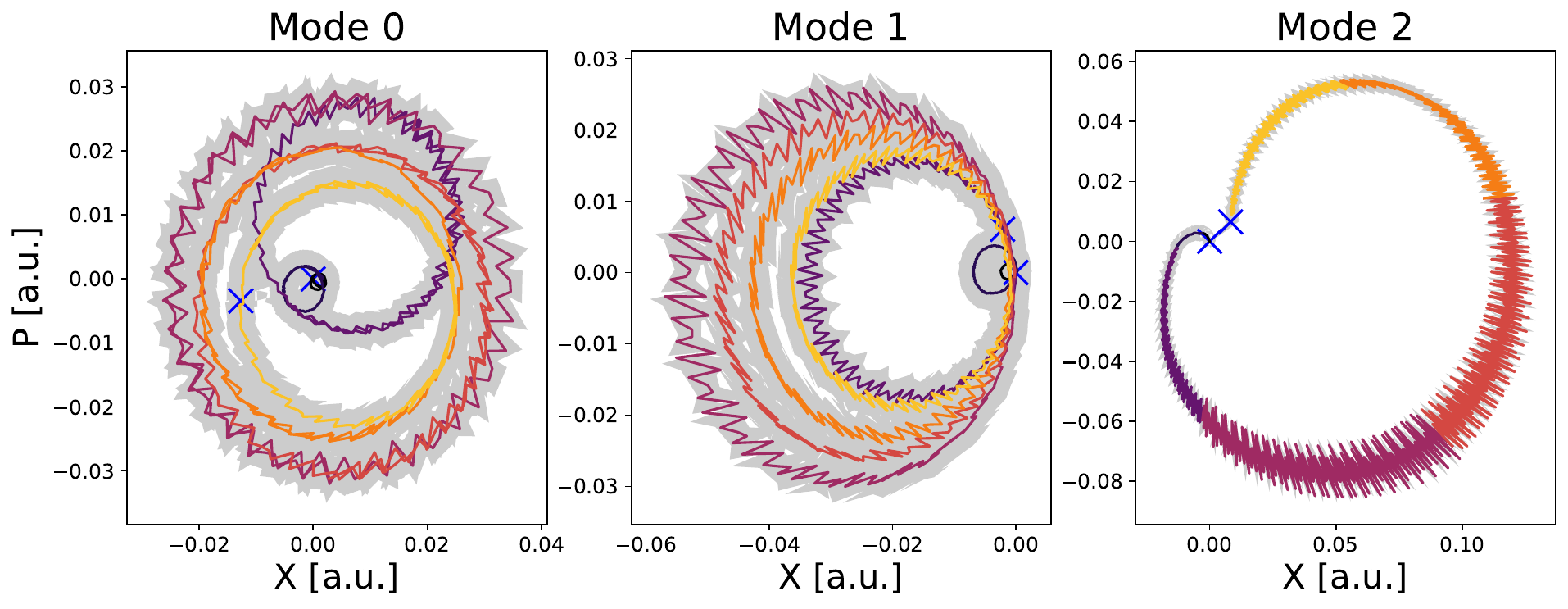}
\caption{Motional \ps\ trajectories of a three-ion chain driven by a reinforcement-learning agent. From left to right, starting at `Mode 0': \com\ mode. 
The quadrature coordinates position $X$ and momentum $P$ are given in arbitrary units (a.u.).
The agent is trained with a nine-segment pulse shape, provided with the current geometric phase and index of observation. 
Each plot shows the average case per motional mode of 1000 independent trajectories (solid line). The shaded 
area represents the actual standard deviation. 
The blue `X' markers represent the starting and ending points of the process. The colour of the trajectory is linked to the piece-wise constant segment of the pulse sequence in Fig.~\ref{fig:pulse_episodic}.}
\label{fig:ps_episodic}
\end{figure}

In Fig.~\ref{fig:ps_episodic} we show the average evolution of the three motional modes over 1000 trajectories exhibiting $\min\mathcal{I} = 46\times 10^{-4}$ and an average $\bar{\mathcal{I}} = 187\times 10^{-4}$.
The learning agent is trained using a seven-segment pulse shape and a neural network with a depth of 256 layers. We notice that, overall, all the motional modes are approximately $10\%$ more excited than the \sota\ solutions. 
This is a consequence of better spectral management, which, in turn, allows the algorithm to navigate the disturbances more efficiently,  as the relative position of the markers demonstrates. 

\subsubsection{\twoq\ employing geometric phase} \label{sec:res:lear:geometric}

\begin{figure}
\centering
    \includegraphics[width=0.6 \linewidth]{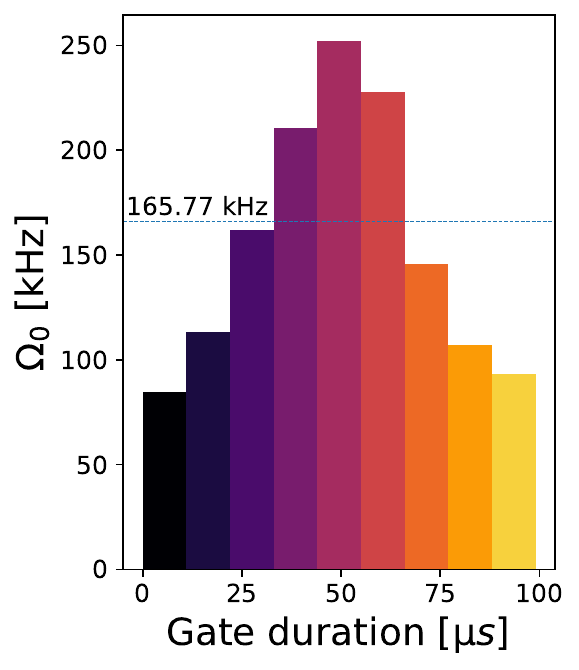}
\caption{Average Nine-segments pulse shape devised employing geometric-phase cost function. The colours are arbitrary and intended to guide the eye over each piecewise constant segment. The blue dashed horizontal line represents the root-mean-square (RMS) value of the pulse sequence. 
The blue dashed horizontal line represents the RMS value of the pulse sequence.}
\label{fig:pulse_current_geo_phase}
\end{figure}

In Fig.~\ref{fig:pulse_current_geo_phase} we can observe the average nin-segment pulse delivering $\min\mathcal{I} = 14\times 10^{-4}$ and $\bar{\mathcal{I}}=85\times 10^{-4}.$ The benefits of employing less power 
are even more present in this scenario, additionally, we observe the trend of a Bell-like shape to some extent, which feature redounds in closing more appropriately the \ps\ trajectories. 
It is worth mentioning that the agent had the means to anticipate when the end of the gate was approaching, as it receives not only an estimation of the current geometric phase but also the observation index~$i$.

\begin{figure}
\centering
   \includegraphics[width=1 \linewidth]{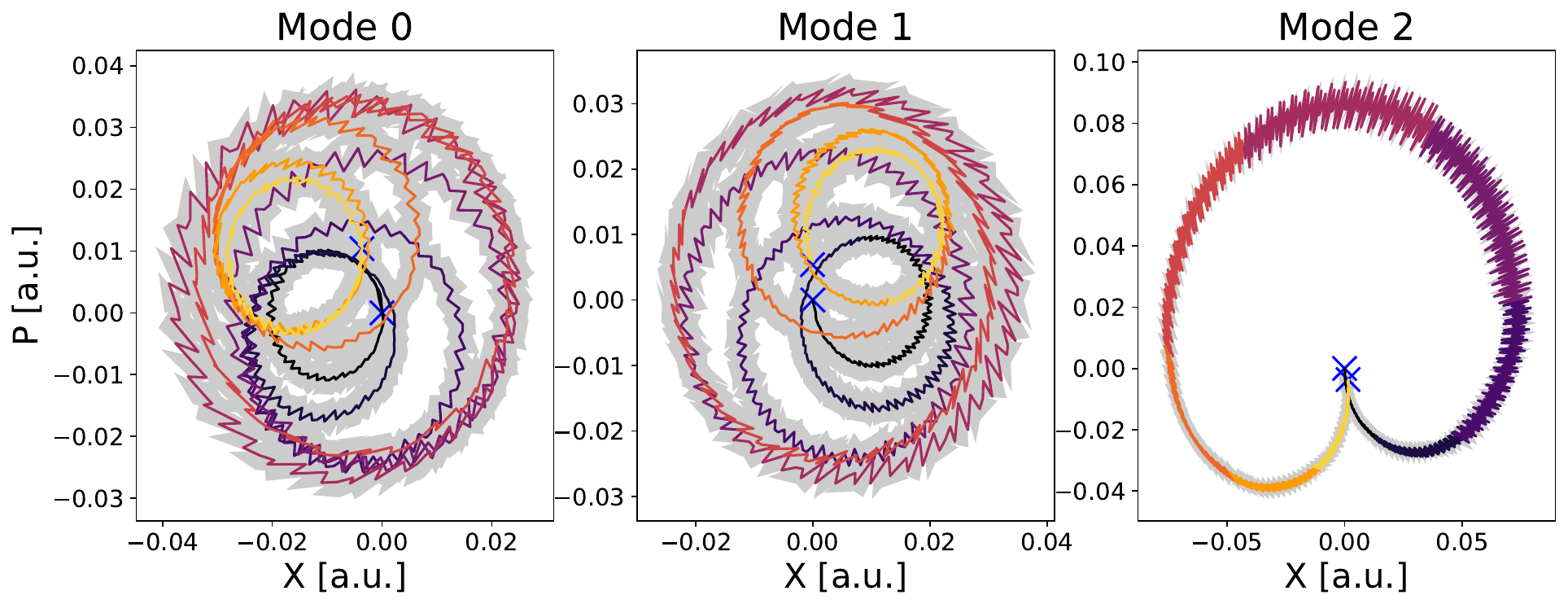}
\caption{Motional \ps\ trajectories of a three-ion chain driven by a reinforcement-learning agent. The quadrature coordinates position $X$ and momentum $P$ are given in arbitrary units (a.u.).
The agent was trained with a nine-segment pulse shape, provided with the current geometric phase and the index of the observation. 
Each motional trajectory shows the average of 1000 independent trajectories (solid lines). 
The agent is trained for a nine-segment piecewise constant pulse shape employing a neural network with 256 layers. The blue `X' markers indicate the starting and ending points of the evolution.
The colour of the trajectory is associated with the piecewise constant segment of the pulse sequence, as shown in Fig.~\ref{fig:pulse_current_geo_phase}. The shaded area represents the actual standard deviation.}
\label{fig:ps_acc}
\end{figure}

In this scenario the agent is only provided with a 
hypothetical geometric phase corresponding exclusively to the current segment. As time progresses the demand to accumulate geometric phase is 
lesser. Such a condition explicitly requires the agent to reduce the amount of energy spent in the last stages of the gate.
In this scenario, we were able to increase the number of pulses and achieve convergence to a solution.
In Fig.~\ref{fig:ps_acc} we show the average of 1000 independent trajectories under our CLPM-scheme.
Note that the X markers overlap for all motional trajectories.

\subsection{Validation procedure}\label{sec:res:validation}

\subsubsection{Test 1: thermal effects}

As the fundamental position measurement block has been experimentally realised, we refer the reader to Refs.~\cite{CAP+21_prl, DHB+22, SRZ05, CAP+21_apl}.
The central idea of this stage is to gather information about the fluorescence signal strength and cooling effects in a three-ion crystal.
This stage also provides preliminary information for determining the optimal operating spot of the spectator ion. Because there is no gate implementation at this stage, we expect a maximum cooling rate $\Gamma_\text{c}\sim \nicefrac{1000}{\sqrt{3}}~\rad$ at $\Delta_\text{L}\sim 10$~MHz.

In general, previous estimations must be considered with caution because they correspond to an effective two-level electronic system. In practice, all possible decay channels from the \ael2P1 state must be considered. 
In particular, the levels \ael2D3 in the context of their Zeeman splitting which can be on the order of $\Delta_\text{L}$, create possible dark states or EIT lines with the \ael2S1 energy level. 
The described scenario can be considerably minimised with larger detunings $\Delta_\text{L}$ because the branching ratio from the \ael2P1 to \ael2D3 is approximately 0.5\%.

\subsubsection{Test 2: induced perturbative errors}

The core idea here is to perturb the gate implementation and reproduce the errors in the \twoq\ infidelity, as shown in Figs.~\ref{fig:spin_error},~\ref{fig:motion_error}, and~\ref{fig:mirror_modes}.
As these three effects are always present, they cannot be completely isolated from one another. This requires the evaluation of gate implementations under different scenarios. 

The first step involves blocking the back mirror and recording an increase of a few parts per ten thousand in the infidelity. 
Such an exercise removes the errors due to the mirror modes coupled back to the spectator ion and the quantum backaction due to the information extraction of the spectator ion's position. 
The effect is illustrated in Fig.~\ref{fig:mirror_modes} where the errors grow with the power of three of the external laser $\Omega_\text{s}$ which is easily verifiable according to Eq.~(\ref{eq:error_mirror}).

The second step consists of gradually setting the detuning closer to the \com\ mode at $\nu_\text{CoM}=4.3807$~MHz. 
Such a large modification of $\Delta_\text{L}$ implies a drop in the concomitant cooling effect, the effects of which are difficult to predict in a real  trap interaction with its particular fingerprint of heating sources.
Therefore, it is necessary to implement the shortest gate duration $\tau_{\rm g}$.
Alternatively, the detuning can be set to coherently drive the carrier modulation of the second-order counter-rotating motional term.

\subsubsection{Validation: error mitigation}

In this last stage, the experimenter effectively closes the control loop by implementing a \twoq, whose learning agent has been previously trained in our simulated environment.
Here, the task at hand involves an experimental validation~\cite{VPS23} of the \twoq\ implementation.
Once the learning agent is trained with the conditions described in the previous stages, the devised control policy should be able to deal with engineered motional heating and drift, reproducing a substantial improvement in the final Bell state compared with the traditional \sota\ method.

\section{Discussion}
We have meticulously developed an innovative control scheme that integrates the position monitoring of a spectator ion and data-driven algorithms which deliver high-quality \twoq. Here, we organise our discussion into four parts. First, we discuss the feasibility of our control scheme. Second, we discuss the effects of ion-position monitoring on \twoq\ implementations. Subsequently, we discuss the results of the data-driven algorithm. Finally, we discuss the proposed roadmap for validating our CLPM scheme.

First, we discuss our control scheme in light of the feasibility conditions stated in Problem~\ref{prob:phys}.
Our findings in~\S\ref{sec:res:feas:latency} indicate that our approach is viable within the current technological landscape and theoretical advancements. 
The signal latency, encompassing both travel time and post-processing time, is significantly shorter than the evolution of the \ps\ trajectories $T_0 \sim 100$~\us, thereby ensuring the Markovian nature of our CLPM-scheme. 
Notably, our calculations reveal that the number of operations necessary to execute the PPO algorithm, which involves computing forward passes through the neural network, is three orders of magnitude lower than the maximum number reported for current state-of-the-art FPGAs~\cite{NS+20, ZW+18}.

In~\S\ref{sec:res:feas:phase_space}, we show that our design allows the reconstruction of motional trajectories. Such reconstruction is not fundamentally constrained but is limited by experimental imperfections and detection efficiencies. 
Importantly, the minimum required $\Gamma_\text{m}$ exceeds $\Gamma_0$ by almost three times, as shown in Fig.~\ref{fig:rabi_rates_mirror} for $\Delta_\text{L}=-10$~MHz.
However, the maximum value of ~$\Omega_{\jmath}$ is maintained below the intensity threshold that optical devices can accommodate, which is equivalent to $\Omega_\text{max} \sim 320$~kHz~\cite{Lin22}.

Second, we examine the outcomes of the three primary processes that contribute to errors in the spin interaction from the position monitoring of the spectator ion. The adverse effects are mathematically represented by Eqs.~(\ref{eq:error_spin}),~(\ref{eq:error_mirror}), and~(\ref{eq:error_motion}) are shown in~\S\ref{sec:res:impact}.

We observe that the spin interaction, as expressed in Eq.~(\ref{eq:error_spin}) of~\S\ref{sec:res:imp:spin}, between the gate ions and the spectator diminishes with the inverse of~$\Delta_\text{L}$.
The reduction in $\mathcal{I}_\text{spin}$ is a consequence of the weaker coupling between the external laser and the spectator ion's dipole moment~$\bm{d}_{\rm eg}$.
While it is advantageous to minimise $\mathcal{I}_\text{spin}$, a trade-off exists with the strength of the scattering rate~$\Gamma_\text{m}$ necessary for reconstructing the \ps\ trajectories.
As shown in Fig.~\ref{fig:rabi_rates_mirror}, the stronger $\Omega_s$ and smaller $\Delta_\text{L}$ the better the collected signal for position monitoring but the higher the errors due to the spin coupling.

Now, we discuss the motion-mediated error $\mathcal{I}_\text{motion}$, which is formally defined in Eq.~(\ref{eq:error_motion}) of~\S\ref{sec:res:imp:motion}.
It is evident from Fig.~\ref{fig:motion_error} that~$\mathcal{I}^{(k)}_\text{motion}$ increases hyperbolically when the laser detuning~$\Delta_\text{L}$ approaches any of the motional mode frequencies~$\bm{\nu}$, as this implies a direct state population transfer. 
Furthermore, $\mathcal{I}_\text{motion}$ increases with the number of motional modes; however, it can be significantly reduced by carefully selecting~$\Delta_\text{L}$.

Now, we consider the mirror-mode-induced error~$\mathcal{I}_\text{mirror}$, as described in Eq.~(\ref{eq:error_mirror}) of~\S\ref{sec:res:imp:mirror_mode}. 
This type of error is the most intriguing among all the associated errors. 
Fundamentally, the nature of this interaction resembles an entangling operation mediated by the collective crystal motion, with the distinction that the driving field~$\bm{E}_\text{m}$ is stochastic by principle. 
Similar to $\mathcal{I}_\text{motion}$, the error due to the mirror modes increases according to the distance $\abs{\left(\Delta_\text{L}-\nu_k \right)}$; however, this distance cannot be arbitrarily large; otherwise, the signal strength for position detection will be unacceptably weak.
In Table~\ref{tab:infidelity_cont_mon}, we demonstrate that the error due to position monitoring does not exceed~$10\times10^{-4}$, accounting for only ~10\% of the overall error budget.

Third, we discuss our results for the \twoq\ implementations employing our CLPM scheme, which are shown in~\S\ref{sec:res:learning}.
It is pertinent to note that the \sota\ method exhibited inferior performance compared to our CLPM-scheme across all test cases, with $\mathcal{I}_{\rm SotA} > 300\times10^{-4}$ for the same error and decoherence sources as those outlined in Table~\ref{tab:infidelity_cont_mon}.
Overall, the dispersion in the \ps\ trajectories was comparable in all instances, with a maximum discrepancy of~0.6\% for both the episodic training and geometric approaches.

Our CLPM-scheme achieves a lower~$\mathcal{I}$ relative to the \sota\ method, with a 33\% improvement for the episodic approach shown in~\S\ref{sec:res:lear:episodic}. 
The episodic approach, which is the simplest of all RL approaches, always offers insight into the versatility of the learning algorithm in navigating plant dynamics.
Despite all sources of noise and decoherence, the algorithm can navigate the intricacies of \twoq\ implementations and the high level of disturbance. 

Our CLPM-scheme employing the geometric-phase-based approach in~\S\ref{sec:res:lear:geometric} achieves the best performance with a~70\% enhancement  with respect to the\sota\ solutions.
The proposed approach employs an agent trained using a 256-layer NN and a nine-segment piecewise constant pulse shape. This pulse shape provides sufficient degrees of freedom to compensate for the errors and decoherence that occur during \twoq\ implementations.
The presence of index $i$ as an input to the learning agent enables it to smoothen and symmetrise the pulse shape, as the algorithm has a `sense of time' during gate evolution.
These two features are desirable for improving the robustness of the \twoq\ control policies and the efficient use of laser energy~\cite{VPS23}.

Notably, the pulse sequences delivered by the learning agent resemble Bell-like shapes and utilise maximum amplitudes that are one-third lower than the value of ~$\Omega_\text{max}$.
In general, less symmetric pulses are less likely to satisfy the closure condition of the motional modes.

Although the pulse shape differs from the \sota\ solutions, the RMS power is moderate, and the control sequences demonstrate greater robustness.
Interestingly, the RMS power of the geometric approach was the lowest among all the scenarios examined, resulting in fewer pulse-dependent errors.
Our agent successfully navigated the \twoq\ requirements and the randomness of the \ps\ trajectories.
The pulse shape delivered by the geometric approach was the closest to a Bell-like shape that the PPO algorithm could achieve.
Less energetic pulse shapes than the \sota\ pulses under the same conditions are desirable because they translate to fewer disturbances owing to power-dependent disturbances, that is, energy-level shifts and photon scattering processes.

In Figs.~\ref{fig:ps_episodic} and~\ref{fig:ps_acc} we observe the good overlap of the `X'-markers implying that the closure condition~\ref{sec:back:control_2qg} was effectively incorporated by the learning agent during training.
The geometric-phase-based approach is more effective than the episodic approach, as shown by the comparison between Figs.~\ref{fig:ps_episodic} and~\ref{fig:ps_acc}.
We observed high modulation and low amplitude owing to the bare electronic transition in all motional modes. Interestingly, the control policy is sufficiently flexible to employ the correct combination of pulse-segment amplitudes to compensate for the carrier modulation.
Our findings reaffirm the initial understanding that the solution search space is highly non-convex and that the \sota\ method provides an inadequate representation of real-world conditions.

We also provide a roadmap for validating the proposed control scheme.
The validation process consists of three independent and relatively straightforward stages. 
The first stage ensures that sympathetic cooling is effectively implemented while assessing the maximum expected signal at the photodetector for the position monitoring.
The second stage verifies that the errors in implementing a \twoq\ are as predicted in \S\ref{sec:res:impact}.
This stage requires the removal of the effect of the mirror modes to observe the errors due to motion when the laser detuning approaches either the first- or second-order motional frequencies.
The final stage of validation consists of pre-training a learning agent with the correct trap parameters. 
In this stage, regardless of how low the final Bell state infidelity is, the entire system should be able to deal with engineered errors, such as motional heating and frequency drift.

There are few aspects to discuss regarding the comparison between the GO algorithm and CLPM-scheme. 
Establishing a solid comparison between both schemes implies a large set of parameter explorations to identify the conditions under which a certain method outruns the other.
This includes the number of ions, detuning and intensity of the Raman lasers, pair of gate ions, and location of the spectator ion, among others. 
Therefore, we decided to compare it with the state-of-the-art method, as performed by the global optimisation algorithm~\cite{VPS23}. 
Moreover, we emphasise that the detuning was not optimised in the CLPM scenario. With a suitable search strategy for gate detuning $\mu$ our approach is likely to generate better results than the GO method. 

Our control approach for \twoq\ with trapped ions is not only theoretically~\cite{HM26, BH25} and experimentally plausible but also delivers low infidelity gates by judiciously dealing with noise and decoherence in real time. 
Contrary to conventional open-loop or model-based methods, our data-driven agent does not require recalculations or search optimisations once the control parameters have drifted.

\section{Conclusions}
In this study, we developed a closed-loop control scheme based on monitoring the spectator ion position for \twoq\ implementations. This control scheme facilitates the compensation of errors and decoherence that occur during the real-time implementation of \twoq. 
We employ quantum trajectory theory to unravel the QME, which subsequently allows us to compute the phase-space trajectories of the motional modes.
A reinforcement-learning agent is utilised to design policies, taking $\left(x_k, p_k \right)$ as inputs to compute the corresponding output for modulating the intensity of the bichromatic Raman beams. 
The learning agent in our scheme is not confined to the proximal policy optimization employed here; potentially more robust and reliable algorithms could be utilised. 
We also discussed the technical feasibility of our scheme, with enhanced prospects owing to recent advancements 
in ion-position monitoring. 
The use of real-time measurements, relying on a learning agent, also provides valuable information for the system's self-calibration and thermometry, thereby reducing the operational overhead during production.

We also provide a roadmap to validate our proposed control scheme in \S\ref{sec:res:validation}.
The validation process consisted of three independent and relatively straightforward stages. 
The first stage ensures that sympathetic cooling is effectively implemented while assessing the maximum expected signal at the photodetector for position monitoring.
The second stage verifies that the errors in implementing a \twoq\ are as predicted in \S\ref{sec:res:impact}. This stage requires removing the effect of the mirror modes to observe the errors due to motion when the laser detuning approaches either the first- or second-order motional frequencies.
The last stage of validation consists of pre-training a learning agent with the correct trap parameters. In this stage, regardless of how low the final Bell state infidelity is, the entire system should be able to deal with engineered errors, such as motional heating and frequency drift.

Further work, such as statistical Kalman filters, can be implemented to reduce the noise in real-time measurements associated with the estimation of both position and momentum.
Although we do not explore these alternatives in this work, the field of quantum filtering offers an intriguing avenue for approaching the reconstruction of momentum during \twoq\ implementations.
We leave the exploration of more efficient alternatives for reconstructing the \ps\ trajectories for future research.

\section*{Credit authorship contribution statement}
\textbf{Eduardo J.~P\'aez:}
Conceptualization,
Methodology,
Investigation,
Visualization,
Writing -- Original Draft.

\textbf{Seyed Shakib Vedaie:}
Conceptualization,
Methodology,
Investigation,
Writing -- Original Draft.

\textbf{Barry C.~Sanders:}
Conceptualization,
Methodology,
Investigation,
Supervision,
Writing -- Review \& Editing.

\section*{Declaration of competing interest}
The authors declare that they have no known competing financial interests or personal relationships that could have influenced the work reported in this paper.

\section*{Data availability}
Data are available upon reasonable request.

\section*{Acknowledgements}
EJP, SSV, and BCS acknowledge the support of the Major Innovation Fund of the Government of Alberta, Canada. SSV would like to thank MITACS for their support. BCS appreciates the financial support from the Natural Sciences and Engineering Research Council of Canada.  

\bibliography{ref_monitoring.bib}

\newpage
\appendix
\onecolumngrid

\section{Errors on \Inf\ due to position-monitoring of a spectator ion}
\label{appendix:A}

We use the truncated Dyson series, if $\mathcal{T}$ is the time ordering operator
\begin{equation}\label{eq:truncated_dyson_appx}
	U = \mathds{1} - \ii \int_0^t \dd t_1 H(t_1)- \frac{1}{2}\mathcal{T}\int_0^t \dd t_1\int_0^t \dd t_2 H(t_1)H(t_2) + O(t^3),
\end{equation}
to determine the evolution $U$.

The stimulation of another internal degree of freedom (spectator ion) generates spin coupling with the target ions. 
This phenomenon has new implications for the ability to generate pure \twoq\ entanglement in this system.

More specifically, the mirror electric field $\bm{E}_\text{m}$ with polarisation $\bm{e}_{\rm m}$ can be expressed as
\begin{equation}
	\bm{E}^{+}_{\text{m}}(x)=\ii \bm{e}_{\rm m}  \int_0^{\infty} \dd \omega \alpha_{\omega} \sin [k(\omega) x] b_{\text{m}}(\omega)\e^{-\ii \omega t}
\end{equation}
where $b_{\text{m}}$ stands for the bosonic field modes obeying the commutation relations $\left[b_{\text{m}}(\omega), b_{\text{m}}^{\dagger}\left(\omega^{\prime}\right)\right] = \delta(\omega - 
\omega')$ and total energy $H_{\text{m}}=\int \dd \omega \omega b_{\text{m}}^{\dagger}(\omega) b_{\text{m}}
(\omega)$ and interact with the atom via the electric dipole moment $\bm{d}_{\rm eg}$, $\alpha_{\omega}$ is a normalisation factor for the electromagnetic field in the mirror mode.

We can proceed similarly with the background modes $b_u$ and the corresponding electric field $\bm{E}_\text{b}$. 
If we denote the emission angle $\theta$ and $u=\cos(\theta)$, in accordance to the dipole radiation pattern with spatial  normalisation $N(u)$, the noisy quantised electric field is
\begin{equation}
	\bm{E}_\text{b}^{+}(x) = \ii \bm{e}_{\rm b}\int_{-1}^{+1} \dd u \sqrt{N(u)} \e^{-\ii u k \hat{x}} b_u(t);
\end{equation}

Let us recap the full Hamiltonian
\begin{align}
	H_{\mathrm{total}}(t) = &- \sum _ {\jmath } ^ { \{r,q\} }  \hbar \Omega _ {\jmath }   \cos\left(\mu t \right) 
	\left[ \overbrace{\sigma^\text{y}_{\jmath}}^A  + \overbrace{\sum _ { k = 0 } ^ { N -1} \eta_{\jmath k} \left( a _ { k } \e^ { -\ii \omega _ { k } t } + a _ { k } ^ { \dagger } \e^{\ii \omega _ { k } t } \right)}^B \right]\sigma^\text{x}_{\jmath} \\
	&+ \left[\overbrace{ \frac{\Omega_{\rm s}}{2}\sigma^{+}_s\e^{\ii(k_{s}\hat{x}_s-\Delta_\text{L} t)}}^C
	-\sigma^{+}_s\e^{-\ii(\Delta_\text{L} t)}  \bm{\mathrm{ d}}_{\text{eg}}\vdot\left(\overbrace{\ii  E_\text{m} b_\text{m}(t)   \sin (k_{\text{eg}} \hat{x}_s)}^D\vb{e}_x + \bm{E}_\text{b}(x)\right) + \text{hc}\right].
\end{align}

\subsection{Gate ions and spectator spin interaction}

Because the terms A and C involved in this interaction commute, the time-ordering operator can be dropped to simplify notation. 
Let us take $A$ from $H(t_1)$ and $B$ from $H(t_2)$ according to Eq.~\ref{eq:truncated_dyson_appx} and using $\beta_{s}=k_s \hat{x}_s$, then we obtain
\begin{align}
	\mathcal{I}_{\rm AC} = &-\frac{\Omega_{\rm s}}{2} \sum_{\jmath}\sigma^\text{y}_{\jmath}\int_0^t \dd t_1\int_0^t \dd t_2\left[ \sigma^{+}_se^{\ii(k_s \hat{x}_s-\Delta_\text{L}t_2)} +  \sigma^{-}_se^{-\ii(k_s \hat{x}_s-\Delta_\text{L}t_2)}\right]\Omega_{\jmath} \cos \left(\mu t_1\right)\nonumber\\
	&=-\frac{\Omega_{\rm s}}{2} \sum_{\jmath}\sigma^\text{y}_{\jmath}\int_0^t \dd t_1\int_0^t \dd t_2 \left[ \sigma^{+}_s(1+ \ii\beta_{k'})e^{-\ii\Delta_\text{L}t_2}+  \sigma^{-}_s(1- \ii\beta_{k'})e^{\ii\Delta_\text{L}t_2}\right]\Omega_{\jmath} \cos \left(\mu t_1\right)\nonumber \\
	&=-\frac{\Omega_{\rm s}}{2} \sum_{\jmath}\sigma^\text{y}_{\jmath}\int_0^t \dd t_1\int_0^t \dd t_2 \left[ \sigma^\text{x}_s\cos(\Delta_\text{L}t_2) + \sigma^\text{y}_s\sin(\Delta_\text{L}t_2)\right.\nonumber\\
	&\left.+\ii\beta_{k'}\left(\sigma^{+}_se^{-\ii\Delta_\text{L}t_2} - \sigma^{-}_se^{\ii\Delta_\text{L}t_2} \right)\right]\Omega_{\jmath} \cos \left(\mu t_1\right).
\end{align}

Now we can incorporate the definition of $\beta_{s}$, neglecting the fast oscillating terms at $\pm(\omega_{ k }+\Delta_\text{L})$ and group terms properly,
\begin{align}
	\mathcal{I}_{\rm AC}=&-\frac{\Omega_{\rm s}}{2} \sum_{\jmath}\sigma^\text{y}_{\jmath}\int_0^t \dd t_1\int_0^t \dd t_2 \left[ \sigma^\text{x}_s\cos(\Delta_\text{L}t_2) + \sigma^\text{y}_s\sin(\Delta_\text{L}t_2)\right] \nonumber \\
	&-\frac{\ii \Omega_{\rm s}}{2} \sum_{\jmath k'}\sigma^\text{y}_{\jmath}\int_0^t \dd t_1\int_0^t \dd t_2  \eta'_{sk'}\left(\da{a}_{k'}\sigma^{+}_s\e^{\ii(\omega_{k'}-\Delta_\text{L})t_2} - a_{k'}\sigma^{-}_se^{-\ii(\omega_{k'}-\Delta_\text{L})t_2}\right)\Omega_{\jmath} \cos \left(\mu t_1\right),
\end{align}
now  we can resolve the double integrals and keeping in mind that $\kappa_{\jmath} = \int^t_0 \dd t_1 \Omega_{\jmath}\cos(\mu t_1)$,
\begin{align} \label{eq:carrier_spin}
	\mathcal{I}_{\rm AC} &=-\frac{\Omega_{\rm s}}{2}\left[\frac{\sigma^\text{x}_s\sin(\Delta_\text{L}t) - \sigma^\text{y}_s\cos(\Delta_\text{L}t)}{\Delta_\text{L}}\right. \nonumber \\
	&\left. +\sum_{k'}\eta'_{sk'}\frac{\sigma^\text{x}_s\left(\da{a}_{k'} \e^{\ii(\omega_{k'}-\Delta_\text{L})t} - a_{k'} \e^{-\ii(\omega_{k'}  -\Delta_\text{L})t}\right) + 	  \ii\sigma^\text{y}_s\left(\da{a}_{k'} \e^{\ii(\omega_{k'}-\Delta_\text{L})t} + a_{k'} \e^{-\ii(\omega_{k'} - \Delta_\text{L})t}\right)}{\omega_{k'}-\Delta_\text{L}} 
	\right] \sum_{\jmath}\sigma^\text{y}_{\jmath}\kappa_{\jmath},
\end{align}
we can use now the short-hand expression $\tilde{x}_s = \da{a}_{k'} \e^{\ii(\nu_k-\Delta_\text{L})t} + a_{k'} \e^{-\ii(\nu_k-\Delta_\text{L})t}$ and the corresponding expression for $\tilde{p}_s$
\begin{equation}
	\frac{\Omega_{\rm s}}{2}\left[\frac{-\sigma^\text{x}_s\sin(\Delta_\text{L}t) + \sigma^\text{y}_s\cos(\Delta_\text{L}t)}{\Delta_\text{L}} +\ii \frac{\sigma^\text{y}_s\tilde{x}_s+\sigma^\text{x}_s \left(\nicefrac{\tilde{p}_s}{\omega_{k'}-\Delta_\text{L}}\right) }{\nu_k-\Delta_\text{L}} \right] \sum_{\jmath}\sigma^\text{y}_{\jmath}\kappa_{\jmath}, 
\end{equation}

The net effect of the purely spin coupling is condensed into the first term. We have to double that value to account for the cross product of the equivalent terms $A$ and $C$ in $H(t_2)$ and $H(t_1)$, respectively, then
\begin{equation}
	\mathcal{I}_\text{spin} = 2\max \mathcal{I}_{\rm AC} =\sum_{\jmath}^{q,r}\frac{\Omega_{\rm s}\Omega_{\jmath}^{\rm max}}{\mu\Delta_\text{L}}\kappa_\text{c},
\end{equation}
which is similar to the spin interaction discussed previously. 
However, the direction of the torque applied to the spin state in the Bloch sphere spins with $\Delta_\text{L} t$ is The factor $\kappa_\text{carrier}$ accounts for the error in the failure to close the carrier term. Let us stress the fact that in any pulse design that considers the carrier term, the idea is to close this term by either pulse shaping or selectively choosing the detuning $\mu$. 

We can observe that the effect of the spectator's motion on the entanglement is expressed in the last term of Eq.~(\ref{eq:carrier_spin})
\begin{equation}
	\frac{\Omega_{\rm s}}{2}\left[\frac{\sigma^\text{x}_s\tilde{x}_s+\sigma^\text{y}_s \left(\nicefrac{\tilde{p}_s}{\omega_{k'}-\Delta_\text{L}}\right) }{\nu_k-\Delta_\text{L}} \right] \sum_{\jmath}\sigma^\text{y}_{\jmath}\kappa_{\jmath},
\end{equation}
whose maximum value is
\begin{equation}
	\mathcal{I}_1 = \frac{\eta'_{sk'}\Omega_{\rm s}\Omega_{\jmath}}{2\mu(\Delta_\text{L}-\nu_k)}.
\end{equation}

This level of entanglement is inferior to the spin term between the target ions if the carrier is not properly closed, as indicated by $\kappa_{\jmath}$.

\subsection{Motion-mediated entanglement}
Because the motion of the spectator is excited by an external laser (terms B and C), this disrupts the collective mode of motion during an ideal \twoq\ gate implementation, and the extent of the disruption depends on the strength and duration of the illumination of the spectator ion.

Proceeding as before, the effect of the  collective mode of motion is,
\begin{align}
	\mathcal{I}_{\rm BC}= -\frac{\Omega_{\rm s}}{2}  \sum_{\jmath}\int_0^t \dd t_1\int_0^t \dd t_2 &\mathcal{T}\left[  \sigma^\text{x}_s\cos(\Delta_\text{L}t_2) + \sigma^\text{y}_s\sin(\Delta_\text{L}t_2)\right.\nonumber\\
	&+\left. \ii\beta_{k'}\left(\sigma^{+}_se^{-\ii\Delta_\text{L}t_2} - \sigma^{-}_se^{\ii\Delta_\text{L}t_2} \right)\right] \sum_k\beta_k\Omega_{\jmath} \cos \left(\mu t_1\right)\sigma^\text{x}_{\jmath},
\end{align}
where we can observe the intricate interaction between the corresponding detuning of the Raman lasers and the external laser on the spectator $\Delta_\text{L}$ and $\mu$.

The first term indicates the interaction between the spectator's off-resonant carrier and the \sdf of the target ions. Since the time-ordering operator is innocuous in this case, we obtain,
\begin{align}
	\mathcal{I}_{\rm BC} = -\frac{\Omega_{\rm s}}{2}& \sum_{\jmath}\int_0^t \dd t_1\int_0^t \dd t_2\mathcal{T}\left[  \sigma^\text{x}_s\cos(\Delta_\text{L}t_2) + \sigma^\text{y}_s\sin(\Delta_\text{L}t_2)\right] \sum_k\beta_k\Omega_{\jmath} \cos \left(\mu t_1\right)\sigma^\text{x}_{\jmath}\nonumber\\
	&=-\frac{\Omega_{\rm s}}{2}\left[\frac{\sigma^\text{x}_s\sin(\Delta_\text{L}t) - \sigma^\text{y}_s\cos(\Delta_\text{L}t)}{\Delta_\text{L}}\right] \sum_{\jmath k}\left(\gamma_{\jmath k} \hat{a}_k^{\dagger}+\gamma_{\jmath k}^* \hat{a}_k\right)\sigma^\text{x}_{\jmath}\nonumber\\
	&=- \frac{\Omega_{\rm s} \sum_{\jmath k}\left(\gamma_{\jmath k} \hat{a}_k^{\dagger}+\gamma_{\jmath k}^* \hat{a}_k\right)\sigma^\text{x}_{\jmath}\sigma_{s}^{\pi/2-\Delta_\text{L}t}}{2\Delta_\text{L}},
\end{align}
where we have defined $\sigma_{s}^{\pi/2-\Delta_\text{L}t} = \sigma^\text{x}_s\sin(\Delta_\text{L}t) - \sigma^\text{y}_s\cos(\Delta_\text{L}t)$

The effect of the spectator's carrier on the motion of the target ions is to generate a fast oscillating  spin direction which is further minimised by the closure condition on $\gamma_{\jmath k}$ according to the pulse-shaping design; hence, we can take this factor to average out.

The second term, in contrast, considers the motion of the spectator and target ions. We can work out this term assuming first that $k = k'$, then we obtain the following integration for the motion-motion cooperation,
\begin{align}
	\mathcal{I}_{\rm BC} =&\frac{\Omega_{\rm s}}{2}\eta'_{sk'}\int_0^t \dd t_1\int_0^t \dd t_2\mathcal{T} \left[ \sigma^\text{x}_s\left({a_{k}^{\dagger}} \e^{\ii(\nu_k-\Delta_\text{L})t_2} - a_{k} \e^{-\ii(\nu_k - \Delta_\text{L})t_2}\right) + \ii\sigma^\text{y}_s\left({a_{k}^{\dagger}} \e^{\ii(\nu_k-\Delta_\text{L})t_2} + a_{k} \e^{-\ii(\nu_k-\Delta_\text{L})t_2}\right)\right. \nonumber\\
	&\left. \times \sum_{\jmath}\Omega_{\jmath}\eta_{\jmath k}\cos(\mu t_1) \left(a_{k}^{\dagger}\e^{\ii\nu_k t_1} + a_{k}\e^{-\ii\nu_k t_1}\right) \right] \sigma^\text{x}_{\jmath},
\end{align}
now we proceed to multiply the harmonic components ignoring the non-conserving terms $a_k a_k$ and $\da{a}_k\da{a}_k$
\begin{align}
	\mathcal{I}_{\rm BC} &=\frac{\Omega_{\rm s}}{2}\eta'_{sk'}\sum_{\jmath}\eta_{\jmath k}\int_0^t \dd t_1\int_0^t \dd t_2\Omega_{\jmath}\mathcal{T}\left[\sigma^\text{x}_s\left({a_{k}^{\dagger}} \e^{\ii(\nu_k-\Delta_\text{L})t_2} a_{k} \e^{-\ii\nu_k t_1}  - a_{k} \e^{-\ii(\nu_k-\Delta_\text{L})t_2} a_{k}^{\dagger} \e^{\ii\nu_k t_1}  \right)\right.\nonumber\\
	&\left. + \ii\sigma^\text{y}_s\left( {a_{k}^{\dagger}} \e^{\ii(\nu_k-\Delta_\text{L})t_2} a_{k} \e^{-\ii\nu_k t_1}  + a_{k} e^{-\ii(\nu_k-\Delta_\text{L})t_2} a_{k}^{\dagger} \e^{\ii\nu_k t_1}  \right)\right] \sigma^\text{x}_{\jmath},
\end{align}
now let us group the harmonic functions at $\nu_k$ with different time labels
\begin{align}
	\mathcal{I}_{\rm BC}&={\frac{\Omega_{\rm s}}{2}\eta'_{sk'}}\sum_{\jmath}\eta_{\jmath k}\int_0^t \dd t_1\int_0^t \dd t_2 \Omega_{\jmath}\mathcal{T}\left[\sigma^\text{x}_s\left({a_{k}^{\dagger}} a_{k} \e^{-\ii\Delta_\text{L}t_2}\e^{\ii\nu_k(t_2-t_1)} - a_{k} a_{k}^{\dagger} \e^{\ii\Delta_\text{L}t_2}\e^{-\ii\nu_k(t_2-t_1)}\right)\right.\nonumber\\
	&\left.+ \ii\sigma^\text{y}_s\left({a_{k}^{\dagger}} a_{k} \e^{-\ii\Delta_\text{L}t_2}\e^{\ii\nu_k(t_2-t_1)} + a_{k}a_{k}^{\dagger} \e^{\ii\Delta_\text{L}t_2}\e^{-\ii\nu_k(t_2-t_1)}\right)\right] \sigma^\text{x}_{\jmath}\nonumber\\
	&={\frac{\Omega_{\rm s}}{2} \eta'_{sk'}}\sum_{\jmath} \left[\sigma^\text{x}_s\left({a_{k}^{\dagger}}a_{k}\varphi_{\jmath s}^* - a_{k} a_{k}^{\dagger}\varphi_{\jmath s}\right) + \ii\sigma^\text{y}_s\left({a_{k}^{\dagger}}a_{k}\varphi_{\jmath s}^* + a_{k}a_{k}^{\dagger}\varphi_{\jmath s}\right)\right]\sigma^\text{x}_{\jmath},
\end{align}
where $\mathcal{T}$ can be immediately resolved since the operators inside commute
\begin{align}
	&\varphi_{\jmath s} =\eta_{\jmath k}\int_0^t \dd t_1\int_0^t \dd t_2\Omega_{\jmath}\mathcal{T}\left[ \e^{\ii\Delta_\text{L}t_2}\e^{-\ii\nu_k(t_2-t_1)}\cos(\mu t_1) \right] \nonumber\\
	&= \eta_{\jmath k}\int_0^t \dd t_1\int_0^t \dd t_2\Omega_{\jmath}\e^{\ii\nu_k t_1}\cos(\mu t_1)\left[ \e^{\ii\Delta_\text{L}t_2}\e^{-\ii\nu_kt_2} \right] 
\end{align}

The upper bound of this integral will have the following form after including a factor of two as done for $\mathcal{I}_{\rm spin}$
\begin{equation}
	\mathcal{I}_\text{motion} = 2\sum_{kk'}\sum_{\jmath}^{q,r}\max \mathcal{I}_{\rm BC} = \sum_{kk'}\sum_{\jmath}^{q,r} \frac{\eta'_{sk'}\eta_{\jmath k} \Omega_{\rm s}\Omega_{\jmath}^{\rm max} \alpha_{\tau}}{\abs{\left(\Delta_\text{L}-\nu_{k'} \right)\left(\mu - \nu_k  \right)}},
\end{equation}
incorporating a damping factor $\alpha_\tau$ due to the closure condition that naturally arises in the expression of $\varphi_{\jmath s}$.

\subsection{Mirror-modes entanglement}
A careful inspection (terms B and D) reveals that the motion of the ion is coupled to the mirror mode via the quantum noise operator $b_\text{m}$ as a consequence of the light reflected from the back mirror. 
In the regime where the light travel is much smaller than the natural decay rate of the dipole transition \ce{^1S_$\nicefrac12$}--\ce{^2P_$\nicefrac12$} transition, with $\Gamma_0 \approx 2\pi 19.6$MHz, we can consider the feedback as a Markovian process and drop out any delay from now onwards.

A similar treatment reveals that the mirror mode coupled back to the ion induces diffusion in its motion, and hence, in the evolution of entanglement.

\begin{align}
	\mathcal{I}_{\rm BD} = \ii \int_0^t \dd t_1\int_0^t \dd t_2 \sin(k_{\text{eg}}\hat{x}_s)\mathcal{T}&\left[d_{\rm eg} E_\text{m} \left( b_\text{m}(t_2)\sigma^+_{\rm s}\e^{\ii\Delta_\text{L} t_2} + b_\text{m}^{\dagger}(t_2)\sigma^-_{\rm s} \e^{-\ii\Delta_\text{L} t_2} \right) \right]\nonumber \\
	&\times \left[ \sum_{\jmath,k}\Omega_{\jmath} \eta_{\jmath k}\cos(\mu t_1) \left(a_{k}^{\dagger} \e^{\ii\nu_kt_1} + a_{k} \e^{-\ii\nu_kt_1}\right)\sigma^\text{x}_{\jmath}\right]
\end{align}

Solving this integral is not easy because it involves a Wiener process and a time-ordering operator. Nonetheless, we can make sensible assumptions for the sole purpose of bounding the error in the worst-case scenario introduced by this interaction, although this is not a rigorous solution.

The first aspect to noticing is that within the Lamb-Dicke regime we can approximate $\sin(k_{\text{eg}}\hat{x}_s) \approx k_{\text{eg}}\hat{x}_s=\sum_{k'}\eta'_{sk'}(a_{k'} \e^{-\ii\nu_{k'} t}+ \da{a_{k'}} \e^{\ii\nu_{k'} t} )$. 
The time scale for the noise operator is linked to the spontaneous emission of the trapped ion, whose linewidth for dipole-allowed transitions is typically $\sim 10$~MHz, so the process is within the same time scale as $\mu$ and~$\Omega_{\rm s}$. 
As in the \ito\ representation the noise operators commute with the system's operators, we can resolve the double integral independently and making $k=k'$, hence obtaining,

\begin{align}
	\mathcal{I}_{\rm BD}^{(k)} = \ii \eta'_{sk'}d_{\rm eg}E_\text{m}\sum_{\jmath}\int_0^t \int_0^t \dd t_1  \dd t_2\Omega_{\jmath} \eta_{\jmath k}&\cos(\mu t_1) \left(a_{k}\e^{-\ii\nu_kt_1} + a_{k}^{\dagger}\e^{\ii\nu_kt_1} \right)\times\nonumber\\
	&\left[a_kb_\text{m}(t_2)\sigma^+_{\rm s}\e^{\ii(\Delta_\text{L} - \nu_k) t_2} + a_{k}^{\dagger}b_\text{m}^{\dagger}(t_2)\sigma^-_{\rm s} \e^{-\ii(\Delta_\text{L} - \nu_k) t_2}\right]\sigma^\text{x}_{\jmath},
\end{align}
which can be further simplified by introducing the Wiener process $\dd W(t) = b_\text{m}(t)\dd t$,
\begin{align}
	\mathcal{I}_{\rm BD}^{(kk')}=\ii 2  \eta'_{sk'}\Omega_\text{m}\sum_{\jmath}\sigma^\text{x}_{\jmath} \eta_{\jmath k}\int_0^t \int_0^t \dd t_1\Omega_{\jmath}&\cos(\mu t_1)\left(a_{k}^{\dagger} \e^{\ii\nu_kt_1} + a_{k} \e^{-\ii\nu_kt_1}\right) \nonumber\\
	&\times \left[a_{k'}\dd W(t_2)\e^{\ii(\Delta_\text{L} - \nu_{k'}) t_2}\sigma^+_{\rm s} + \dd W^{\dagger}(t_2)a^{\dagger}_{k'}\sigma^-_{\rm s} \e^{-\ii(\Delta_\text{L}-\nu_{k'}) t_2} \sigma^-_{\rm s}\right],
\end{align}
where we have used $\Omega_\text{m} = d_{\rm eg}E_\text{m}/2$.

Meanwhile, the Wiener process is a Gaussian-amplitude-distributed density function, and we can expect that the  \ito\ integral becomes a normal distribution with a variance proportional to the integration time $t$~\cite{GZ04}. The \ito\ integral is
\begin{equation}
	\int_0^t f(s) d W_s=\lim _{n \rightarrow \infty} \sum_{k=0}^{2^n-1} f\left(t_{n, k}\right)\left(W_{t_{n, k+1}}-W_{t_{n, k}}\right),    
\end{equation}
where the harmonic $\e^{\ii\Delta_\text{L} t_2}$ and the Wiener increment $\dd W(t)$ are independent of each other. If we let $t_{n, k}=\frac{k}{2^n} t$, therefore,
\begin{align}
	\sum_{k=0}^{2^n-1} f\left(t_{n, k}\right)\left(W_{t_{n, k+1}}-W_{t_{n, k}}\right) \sim & \mathcal{N}\left(0, \sum_{k=0}^{2^n-1} \operatorname{Var}\left(f\left(t_{n, k}\right)\left(W_{t_{n, k+1}}-W_{t_{n, k}}\right)\right)\right)\\
	&= \mathcal{N}\left(0, \sum_{k=0}^{2^n-1} f\left(t_{n, k}\right)^2 2^{-n} t\right),
\end{align}
identifying $\lim _{n \rightarrow \infty} \sum_{k=0}^{2^n-1} f\left(t_{n, k}\right)^2 2^{-n} t=\int_0^t\left(f(s)\right)^2 \dd s$, then
\begin{equation}
	\int_0^{\tau_\text{g}}\dd W(t_2)\e^{\ii(\Delta_\text{L}-\nu_k) t_2} = \mathcal{N}\left(0,\int_0^{\tau_\text{g}} \dd t_2\e^{2\ii(\Delta_\text{L}-\nu_k) t_2} \right)= \mathcal{N}_{\tau},
\end{equation}
whose variance is $\nicefrac{1}{2(\Delta_\text{L}-\nu_k)}$.

We can notice the interplay between all elements in the experimental layout, the intensity of the laser coupling on the spectator in $\Omega_\text{m}$, the participation of the motional modes in the Lamb-Dicke parameters, and the mean occupation number.

After removing the fast oscillating terms $\pm(\Delta_\text{L}+\nu_{k'})$, the resultant expression for the partial error is,
\begin{equation}
	\mathcal{I}_{\rm BD} = 2 \ii \Omega_\text{m}\sum_{\jmath} \sum_{kk'}\eta'_{sk'}\left(\mathcal{N}_{\tau} a_{k'}\sigma^+_{\rm s} + \mathcal{N}_{\tau}^* a^{\dagger}_{k'}\sigma^-_{\rm s} \right)\times\left(\alpha_{\jmath k}a_{k}^{\dagger} - \alpha^*_{\jmath k}a_{k}  \right)\sigma^\text{x}_{\jmath},
\end{equation}
where the first parenthesis is the Jaynes-Cumming interaction driven by $\mathcal{N}_\tau$, and the second parenthesis is the dynamical phase $\alpha_{\jmath k}(t)$.

To bound the error due to the red-sideband excitation related to the first parenthesis, we'll replace that term by probability of populating the higher energy level
\begin{equation}
	\rho_{\rm e e}=\frac{s / 2}{1+s+\left(2 \delta_{\mathrm{eff}} / \Gamma\right)^2},
\end{equation}
where $s=2\Omega^2 / \Gamma^2$. Considering $s$ is a very small parameter, we can approximate $\rho_{\rm e e} \sim \Omega^2_{\rm m}/4(\Delta_\text{L} - \nu_k)^2$.

We can now write the total error owing to the electromagnetic mirror-mode, replacing $\mathcal{N}_{\tau}$ by its standard deviation
\begin{equation}
	\mathcal{I}_\text{mirror} = 2\sum_{kk'}\sum_{\jmath}\max \mathcal{I}_{\rm BD}^{(kk')} =\sum_{kk'} \sum_{\jmath}\frac{\eta'_{sk'}\eta_{\jmath k}\Omega^3_\text{m} \Omega_{\jmath}^{\rm max} \alpha_{\tau}}{\abs{\mu-\nu_k}\sqrt{\left(\Delta_\text{L}-\nu_{k'}\right)^5}}.
\end{equation}
\end{document}